\documentclass[a4paper,usenatbib]{mnras}
\usepackage{graphicx}
\usepackage{bm}
\usepackage{amssymb}
\title[3D PIC simulations of ABC fields]{Three-dimensional kinetic simulations of relativistic magnetostatic equilibria}

\author[K. Nalewajko]{
Krzysztof Nalewajko
\\
Nicolaus Copernicus Astronomical Center, Polish Academy of Sciences, Bartycka 18, 00-716 Warsaw, Poland
\\
\tt{knalew@camk.edu.pl}
}

\begin{document}

\maketitle

\begin{abstract}
We present the results of three-dimensional kinetic particle-in-cell (PIC) simulations of isotropic periodic relativistically magnetized pair-plasma equilibria known as the ABC fields.  We performed several simulations for initial wavenumbers $k_{\rm ini}=2$ or $k_{\rm ini}=4$, different efficiencies of radiative cooling (including radiation reaction from synchrotron and inverse Compton processes), and different mean magnetization values.  These equilibria evolve by means of ideal coalescence instability, the saturation of which generates \emph{ab initio} localized kinetically-thin current layers --- sites of magnetic reconnection and non-thermal particle acceleration --- eventually relaxing to a state of lower magnetic energy at conserved total magnetic helicity.  We demonstrate that magnetic relaxation involves in addition localized collapses of magnetic minima and bulk mergers of current layer pairs, which represents a novel scenario of spontaneous magnetic dissipation with application to the rapid gamma-ray flares of blazars and of the Crab Nebula.  Particle acceleration under strong radiative losses leads to formation of power-law indices $N(\gamma)\propto\gamma^{-p}$ up to $p\simeq-2.3$ at mean hot magnetization values of $\left<\sigma_{\rm hot}\right>\sim{6}$.  Individual energetic particles can be accelerated within one light-crossing time by electric fields that are largely perpendicular to the local magnetic fields.  The energetic particles are highly anisotropic due to the kinetic beaming effect, implying complex patterns of rapid variability.  A significant fraction of the initial total energy can be radiated away in the overall process of magnetoluminescence.
\end{abstract}


\section{Introduction}
\label{sec_intro}

There are several examples of extreme astrophysical environments where magnetic fields are thought to dominate the local energy density including the rest-mass density. Accreting black holes threaded by large net magnetic fluxes are the launching sites of relativistic jets found in some active galactic nuclei (AGN) and stellar X-ray binaries \citep{Bla77,Beg84}. The relativistic (apparently superluminal) motions of jet elements are best explained by conversion of relativistic magnetization $\sigma = B^2/(4\pi w) > 1$ to relativistic 4-velocity $u = \Gamma\beta > 1$ \citep{Li92,Kom07}. In the pulsar wind nebulae, a relativistically magnetized (striped) wind converts into a weakly magnetized fluid subject to a strong termination shock \citep{Cor90,Lyub01,Zra17}.

In addition to such bulk-energy conversions, relativistically magnetized plasmas are potentially powerful particle accelerators. Localized inversions of magnetic line topology, allowed by alternating large-scale currents or current-driven instabilities, create conditions for the process of relativistic magnetic reconnection that is able to sustain strong electric fields that accelerate particles directly or stochastically \citep{Zen01,Hos12,Sir14}. If in turn, the accelerated particles are subject to strong radiative losses, a substantial fraction of the magnetic energy can be converted into non-thermal radiation in the overall process dubbed \emph{magnetoluminescence} \citep{Bla15}.

The idea of magnetoluminescence was invoked to explain extremely rapid and luminous flares of gamma-ray radiation observed in certain blazars \citep{Ack16}, but also to explain incoherent gamma-ray flares observed in the Crab Pulsar Wind Nebula (PWN) \citep{Tav11,Abd11}, and with potential application to magnetar outbursts, gamma-ray bursts and similar phenomena. The characteristic feature of magnetoluminescence is rapid (of the order of light-crossing timescale) and efficient conversion of initially dominant magnetic energy into radiation.

Magnetoluminescence can be demonstrated directly by kinetic numerical simulations employing the particle-in-cell (PIC) algorithm that feature acceleration of particles at the reconnection sites and their subsequent radiative losses. The first numerical experiments motivated primarily by the Crab flares, used relativistic current layers of the Harris type as initial condition for relativistic reconnection with radiation reaction to the synchrotron (SYN) process \citep{Cer13,Cer14}.

An alternative initial condition has been considered that contains no thin (on the kinetic scales) current layers -- these are so-called \emph{Arnold--Beltrami--Childress} (ABC) fields \citep{Dom86,Eas15}.
ABC fields consist of monochromatic magnetostatic waves supported by smoothly distributed currents.
Except for the lowest-order case they are subject to coalescence instability. During the linear stage of this instability one can observe the formation of kinetically thin current layers. The structure of these layers is different from that of the Harris layer -- even in the pair plasma composition a separation of transverse scales is found, with plasma density compressing on the skin-depth scale, and non-ideal electric field forming on the broader gyration scale of the most energetic electrons \citep{Nal16}.

Simulations of ABC fields have been performed with relativistic MHD and force-free algorithms in both 2D and 3D \citep{Eas15,ZraEas16}. A major difference between these results is that 3D ABC fields reach the ground (Taylor) state determined by global helicity conservation \citep{Tay74}, while 2D ABC fields do not reach the Taylor state due to additional topological constraints imposed by plane symmetry \citep{ZraEas16}. On the other hand, kinetic (PIC) simulations of ABC fields have so far only been performed in 2D \citep{Nal16,Yua16,Lyu17}.

In this work, we present the results of the first 3D PIC simulations of relativistic ABC fields. Several cases are considered, including inefficient or efficient radiative cooling, radiation reaction due to synchrotron and inverse Compton processes. We investigate the global efficiency of energy conversions, detailed mechanism of magnetic dissipation, non-thermal particle acceleration and anisotropy.
Selected results concerning the radiation spectra and light curves have been presented in \cite{Nal18}.

The plan of this work is the following: Section \ref{sec_setup} describes the numerical setup and simulation parameters; Section \ref{sec_res} presents the simulation results, including the composition and evolution of total system energy (Section \ref{sec_res_totene}), detailed morphological description of the mechanism of magnetic dissipation (Section \ref{sec_res_maps}), statistics of the volume distribution of magnetic and electric fields (Section \ref{sec_res_voldist}), particle momentum distribution (Section \ref{sec_res_partacc}), particle angular distribution (Section \ref{sec_res_partang}), and detailed behavior of individual tracked energetic particles (Section \ref{sec_res_partind}); Section \ref{sec_disc} contains the discussion, and Section \ref{sec_conc} the conclusions.

\section{Numerical setup}
\label{sec_setup}

We performed a set of three-dimensional kinetic particle-in-cell (PIC) simulations on Cartesian numerical grids with periodic boundaries, using a modified version of the public explicit PIC numerical code {\tt Zeltron} \citep{Cer13}. Modifications from the public version include implementation of a charge-conserving current deposition scheme \citep{Esi01} and of the Vay particle pusher \citep{Vay08}.

The initial configuration of our simulations is a three-dimensional (ABC) magnetic field structure defined as \citep{Eas15}:
\begin{eqnarray}
\label{eq_abc3d}
B_x(x,y,z) &=& B_0\left[\sin(\alpha_k z)+\cos(\alpha_k y)\right]\,,
\nonumber
\\
B_y(x,y,z) &=& B_0\left[\sin(\alpha_k x)+\cos(\alpha_k z)\right]\,,
\\
\nonumber
B_z(x,y,z) &=& B_0\left[\sin(\alpha_k y)+\cos(\alpha_k x)\right]\,,
\end{eqnarray}
where $\alpha_k = 2\pi k/L$ is a constant for wavenumber $k = k_{\rm ini}$ and $L$ is the linear size of the simulation domain: $x,y,z\in[0:L]$.
This configuration satisfies the Beltrami condition $\bm\nabla\times\bm{B} = \alpha_k\bm{B}$, hence it can be simply related to a magnetic vector potential $\bm{A} = \bm{B}/\alpha_k$ and magnetic helicity $H = \bm{A}\cdot\bm{B} = B^2/\alpha_k$.
The case of $k_{\rm ini} = 1$ corresponds to the lowest-energy stable ground state.
The case of $k_{\rm ini} = 2$ is the lowest-energy unstable configuration that we are investigating in this work.
Although in 2D there exists a smaller periodic unstable configuration obtained by rotating the coordinate system \citep{Nal16,Yua16}, we have not identified a smaller periodic unstable setup in 3D.
Note that the magnetic field strength is not uniform:
\begin{eqnarray}
\left(\frac{B}{B_0}\right)^2 &=&
3+2\sin(\alpha_k z)\cos(\alpha_k y)+2\sin(\alpha_k x)\cos(\alpha_k z)
\nonumber\\&&
+2\sin(\alpha_k y)\cos(\alpha_k x)\,,
\end{eqnarray}
hence the mean magnetic energy density is $\left<U_{\rm B}\right> \equiv \left<B^2\right>/(8\pi) = 3B_0^2/(8\pi) \equiv 3U_0$,\footnote{Unless stated otherwise, $\left<\cdot\right>$ denotes average over the whole simulation domain volume or over all particles at fixed time.} and the maximum magnetic field strength is $B_{\rm max} = \sqrt{6}B_0$.
In order to obtain an equilibrium satisfying the Ampere's law (i.e., to suppress the displacement currents), current density $\bm{j} = (k_{\rm ini}c/2L)\bm{B}$ is provided by a population of relativistic particles characterized
by the dipole moment of the local angular distribution of particle momenta $a_1 = (B/B_0)\tilde{a}_1$, where $\tilde{a}_1 \le B_0/B_{\rm max} = 1/\sqrt{6}$ is a constant (with opposite dipole vectors for electrons and positrons in order to cancel out their bulk velocities),
and by uniform number density (including both electrons and positrons) $n = 3k_{\rm ini}B_0/(2e\tilde{a}_1L)$.

Since ABC fields are characterized by uniform magnetic pressure, the initial pressure equilibrium can be satisfied with uniform gas pressure of arbitrary value.
The non-uniform dipole moment of the particle angular distribution does not contribute to the gas pressure.
We set the initial particle energy distribution to the Maxwell-J\"{u}ttner distribution $f(\gamma) = \gamma u/[\Theta\,K_2(1/\Theta)\exp(\gamma/\Theta)]$ with relativistic temperature $\Theta = k_{\rm B}T/(m_{\rm e}c^2)$, where $\gamma = (1-\beta^2)^{-1/2} = E/(mc^2)$ is the particle Lorentz factor or dimensionless energy, $\bm{u} = \gamma\bm\beta = \bm{p}/(m_{\rm e}c)$ is the particle 4-velocity or dimensionless momentum, $\bm\beta = \bm{v}/c$ is the particle dimensionless velocity, and $K_n(x)$ is the modified Bessel function of the second kind.

The corresponding mean hot magnetization value is given by $\left<\sigma_{\rm hot}\right> = \left<B^2\right>/(4\pi w) \simeq (3/2)(U_{\rm B}/U_{\rm e})$, where $w \simeq 4\Theta nm_{\rm e}c^2 = (4/3)U_{\rm e}$ is the ultra-relativistic specific enthalpy and $U_{\rm e}$ is the energy density of the electron-positron gas:
\begin{equation}
\label{eq_sigma_hot}
\left<\sigma_{\rm hot}\right> \simeq \frac{\tilde{a}_1}{8\pi}\left(\frac{L/k_{\rm ini}}{\rho_0}\right)
\,,
\end{equation}
where $\rho_0 = \Theta m_{\rm e}c^2/(eB_0)$ is the nominal gyroradius.
The characteristic property of ABC fields is that magnetization scales linearly with the scale separation between the magnetic field coherence scale (of order $L/k_{\rm ini}$) and the kinetic gyration scale $\rho_0$ \citep{Nal16}. This is because a minimum particle number density is required in order to support the smoothly distributed current density.

We performed five large PIC simulations of 3D ABC fields, the parameters of which are summarized in Table \ref{tab_param}.
Four of these simulations were performed for the $k_{\rm ini} = 2$ configuration, those are denoted as `k2\_*', one simulation was performed for the $k_{\rm ini} = 4$ configuration (k4\_T5\_1152P).
In three simulations, the initial particle temperature was set at $\Theta = 10^5$ (*\_T5\_*), in two other simulations the relativistic temperature was set at $\Theta = 10^6$ (*\_T6*).
One simulation was performed on the \emph{Mira} supercomputer on the numerical grid with $N_x = N_y = N_z = 1024$ cells (*\_M1024), and four simulations were performed on the \emph{Prometheus} supercomputer on the numerical grid with $N_x = N_y = N_z = 1152$ cells (*\_P1152).

All simulations were performed with radiation reaction due to the synchrotron (SYN) process, while simulation k2\_T6ic\_1152P includes in addition radiation reaction to the inverse Compton (IC) process, assuming a uniform isotropic soft radiation field characterized by energy density $U_{\rm ext} = U_0$ and photon energy $E_{\rm ext} = 0.01\;{\rm eV}$. See \cite{Nal18} for the actual radiation reaction formulae applied in our PIC simulations.

The reason for adopting ultra-relativistic particle temperatures $\Theta \sim 10^5 - 10^6$ is to obtain efficient radiative cooling due to synchrotron (and IC) mechanisms. The nominal synchrotron cooling length is given by:
\begin{equation}
\label{eq_lcool}
l_{\rm cool} = \frac{\left<\gamma\right>}{\left<|{\rm d}\gamma/c{\rm d}t|\right>} = 
\frac{\left<\gamma\right>}{\left<\gamma^2\right>}\frac{3m_{\rm e}c^2}{4\sigma_TU_{\rm cool}}
\simeq
\frac{(\pi/2)e}{\sigma_T\Theta^2B_0}\rho_0
\end{equation}
where $U_{\rm cool} = \left<U_B\right> = 3U_0$ is the effective synchrotron cooling energy density, and we used the following statistics of the Maxwell-J\"{u}ttner distribution in the limit of $\Theta \gg 1$: $\left<\gamma\right> \simeq 3\Theta$ and $\left<\gamma^2\right> \simeq 12\Theta^2$. In the case of $\Theta = 10^5$, we obtain $l_{\rm cool}/\rho_0 \simeq 1.1\times 10^5$ (we will refer to it as slow cooling), and in the case of $\Theta = 10^6$, we obtain $l_{\rm cool}/\rho_0 \simeq 1100$ (fast cooling).

The common parameter settings include: the nominal magnetic field strength $B_0 = 1\;{\rm G}$, and the initial number of particles per cell is ${\rm PPC} = 16$ (including both electrons and positrons).
While all simulations are performed on numerical grids of similar size, the physical size of simulation domain is in the range $L/\rho_0 \sim 400 - 900$, as reported in Table \ref{tab_param}.
The different physical sizes correspond to different numerical resolutions $\Delta x^i/\rho_0 = (L/\rho_0)/N_c \sim 0.39 - 0.78$, but also to different mean magnetization values.
One should also note that we used two different values for normalization of the particle dipole moment: a maximum value $\tilde{a}_1 = 0.4$ for simulations k2\_T5\_1024M and k2\_T5\_1152P, and a reduced value $\tilde{a}_1 = 0.2$ for the remaining simulations in order to relax the local anisotropy of particles and the amount of current density per particle density unit $\bm{j}/n$.
Hence, the initial mean hot magnetization values were in the range $\sigma_{\rm hot,ini} \simeq 0.9-7.2$ (see Table \ref{tab_param}).
For the three k2\_*\_1152P simulations, we traded numerical resolution for a more efficient non-thermal particle acceleration allowed by the higher magnetization value.
However, for simulations with reduced value of $\tilde{a}_1$, the effective magnetization value was further reduced.

In the simulations k2\_T5\_1024M and k4\_T5\_1152P, the coalescence instability triggered spontaneously after $ct/L \simeq 4$ due to initial random noise in the particle angular distribution. For the remaining 3 simulations (k2\_T5\_1152P, k2\_T6\_1152P, k2\_T6ic\_1152P), in order to speed-up the development of instability and save the computational cost, we applied a small perturbation to the initial magnetic field distribution, identified as the dominant instability mode by performing Fourier analysis of the k2\_T5\_1024M simulation results:
\begin{eqnarray}
\label{eq_Bpert}
B_{1,x} &=& B_1\left[
-\cos\left(\frac{\alpha_k}{2}(x+y)\right)
-\sin\left(\frac{\alpha_k}{2}(x+y)\right)\right]\,,
\nonumber
\\
B_{1,y} &=& B_1\left[
\cos\left(\frac{\alpha_k}{2}(x+y)\right)
+\sin\left(\frac{\alpha_k}{2}(x+y)\right)\right]\,,
\\
\nonumber
B_{1,z} &=& \sqrt{2}\;B_1\left[
\cos\left(\frac{\alpha_k}{2}(x+y)\right)
-\sin\left(\frac{\alpha_k}{2}(x+y)\right)\right]\,,
\end{eqnarray}
with the perturbation amplitude set at $B_1 = 0.01B_0$. In the Appendix \ref{app1}, we derive analytically a dispersion relation for this particular mode.

\section{Results}
\label{sec_res}

\subsection{Total energy and magnetic helicity}
\label{sec_res_totene}

\begin{figure*}
\includegraphics[width=\textwidth]{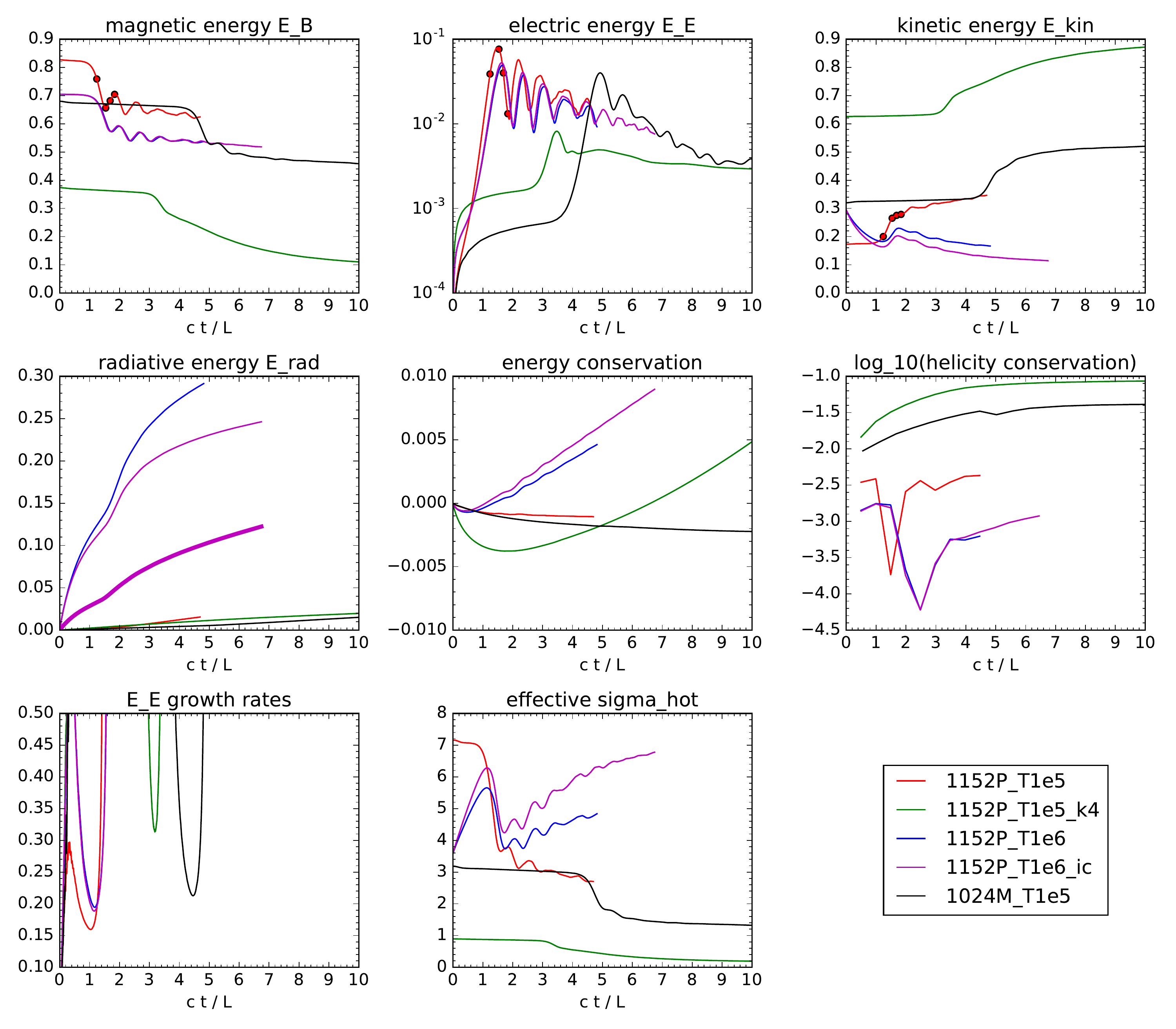}
\caption{The top panels show components of the total energy (magnetic $\mathcal{E}_{\rm B}$, electric $\mathcal{E}_{\rm E}$, and kinetic $\mathcal{E}_{\rm kin}$) integrated over the simulation domain as functions of simulation time $ct/L$ compared for our 5 simulations. Kinetic energy is presented jointly for positrons and electrons. The circles indicate energy levels for the 4 simulation epochs of simulation k2\_T5\_1152P presented in Figure \ref{fig_tot_maps}. The middle left panel shows the cumulative energy radiated away by all particles, including the synchrotron radiation (thin lines) and IC radiation (thick line for simulation k2\_T6ic\_1152P). The center panel shows conservation accuracy of the total energy $\mathcal{E}_{\rm tot}/\mathcal{E}_{\rm tot,0}-1$. The middle right panel shows the conservation accuracy of the total magnetic helicity $\mathcal{H}/\mathcal{H}_0-1$. The bottom left panel shows the growth rate of electric energy defined as $\tau_{\rm E} = [{\rm d}\log E_{\rm E}/{\rm d}(ct/L)]^{-1}$. The bottom center panel shows the effective mean hot magnetization $\left<\sigma_{\rm hot}\right> = 3U_{\rm B}/(2U_{\rm e})$.}
\label{fig_tot_ene}
\end{figure*}

Figure \ref{fig_tot_ene} presents the time evolution of components of the total system energy for the 5 simulations. The initial configurations are in static equilibrium, which is evident from initially constant magnetic energy and very low electric energy. The initial magnetic energy fraction ranges from 37\% for simulation k4\_T5\_1152P to 83\% for simulation k2\_T5\_1152P.

Coalescence instability can be seen as rapid exponential growth of the electric energy, and its saturation leads to a rapid decrease of the magnetic energy, which is a signature of magnetic dissipation. For simulations initiated without seed perturbation of magnetic field (k2\_T5\_1024M and k4\_T5\_1152P), it takes about 3-5 light-crossing times for the coalescence instability to saturate (evidenced by the first peak of electric energy). On the other hand, for simulations initiated with seed magnetic perturbation described by Eq. (\ref{eq_Bpert}), it only takes 1.5 light-crossing times to saturation.

We define the coalescence instability growth rate $\tau_{\rm E}$ as the e-folding time scale of the electric energy $\left<E^2\right> \propto \exp(ct/L\tau_{\rm E})$. The bottom left panel of Figure \ref{fig_tot_ene} shows the minima of local growth rates. The minimum values for each simulation $\tau_{\rm E,min}$ are listed in Table \ref{tab_param}. The most rapid instability growth $\tau_{\rm E,min} \simeq 0.16$ is recorded for simulation k2\_T5\_1152P, and the slowest instability growth $\tau_{\rm E,min} \simeq 0.32$ is recorded for simulation k4\_T5\_1152P. These also happen to be simulations with the highest and the lowest values, respectively, of mean hot magnetization. A trend of faster instability growth rate for higher magnetization has been previously established for the case of 2D ABC fields \citep{Nal16}. It should be noted that the fastest recorded growth rate is still slower than the limiting value of $\tau_{\rm E} = 0.129$ measured in simulations performed with the force-free electrodynamics algorithm corresponding to the limit of $\sigma \to \infty$ \citep{Eas15}.

The peak value of the total electric energy compared with the initial total magnetic energy is a measure of effective global reconnection rate $\beta_{\rm rec} \equiv [\left<E^2\right>_{\rm peak} / \left<B^2\right>_{\rm ini}]^{1/2}$. With such definition, we obtain values in the range $\beta_{\rm rec} \sim 0.24 - 0.31$ for the case of $k = 2$, the highest one for simulation k2\_T5\_1152P, and significantly lower $\beta_{\rm rec} \simeq 0.15$ for the case of $k = 4$.

We also evaluate the efficiency of magnetic dissipation as $f_{\rm B} = 1-\left<B^2\right>_{\rm fin}/\left<B^2\right>_{\rm ini}$, comparing the initial and final total magnetic energies. The values of $f_{\rm B}$ for each simulation are reported in Table \ref{tab_param}, we find them in the range $f_{\rm B} \sim 0.25 - 0.32$ for the case of $k = 2$, and significantly higher $f_{\rm B} \simeq 0.7$ for the case of $k = 4$.

The bottom center panel of Figure \ref{fig_tot_ene} shows the time evolution of the mean hot magnetization value evaluated from the ratio of magnetic to kinetic energies $\sigma_{\rm hot} \simeq (3/2)(U_{\rm B}/U_{\rm e})$. The initial values at $t=0$ are reported in Table \ref{tab_param} and are consistent with Eq. (\ref{eq_sigma_hot}). In the case of slow cooling ($\Theta = 10^5$), magnetization values decrease during magnetic dissipation by factors $\gtrsim k_{\rm ini}$. In the case of fast cooling ($\Theta = 10^6$), magnetization values increase systematically due to radiative losses of gas enthalpy, reaching the peak value of $\sigma_{\rm hot} \simeq 6$ at $ct/L = 1$ before the main phase of magnetic dissipation, then decreasing to $\sigma_{\rm hot} \simeq 4$ at $ct/L = 2$, and then increasing again due to further radiative losses.

The center panel of Figure \ref{fig_tot_ene} shows the conservation accuracy for total energy, including the total energy radiated due to synchrotron and IC processes. Energy conservation is generally better than $\sim 1\%$, with the highest accuracy ($\sim 0.1\%$ by $ct/L \sim 5$) achieved in simulation k2\_T5\_1152P. Strong radiative losses for $\Theta = 10^6$ have a detrimental effect on the energy conservation.

The middle right panel of Figure \ref{fig_tot_ene} shows the conservation accuracy for total magnetic helicity $\left<H\right>$. For simulations with $\Theta = 10^6$ (fast-cooling regime), total magnetic helicity is conserved at the level of $\sim 0.1\%$. However, for the $k = 4$ simulation, the accuracy is only within $\sim 10\%$. It appears that helicity conservation depends significantly on the effective magnetization (enhanced additionally in the case of $\Theta = 10^6$ due to radiative losses of gas enthalpy).

\subsection{3D and 2D maps}
\label{sec_res_maps}

We will describe here a qualitative picture of our simulations around the critical moment of saturation of coalescence instability and the associated magnetic dissipation.
Figure \ref{fig_tot_maps} shows snapshots from the simulation k2\_T5\_1152P for the 4 moments indicated in Figure \ref{fig_tot_ene}. This period of time brackets the first peak of total electric energy, corresponding to the first minimum of total magnetic energy, and to the major particle heating phase.
We present the 3D volume rendering of magnetic field strength $B$ and of the non-ideal field scalar $\bm{E}\cdot\bm{B}$.
We then focus on a single 2D surface at $z = 0$, which is representative for the overall simulation domain.
We further focus on a particular Patch A on the $z = 0$ surface, defined by $750 < x/\rho_0 < 820$ and $500 < y/\rho_0 < 850$, for which we extract 1D profiles of various plasma parameters that are presented in Figure \ref{fig_prof1d}.

The initial volume distribution of magnetic field strength involves a regular network of magnetic minima, around which the magnetic field strength has local minima.
By $ct/L = 1.24$, some of these magnetic minima develop localized magnetic reconnection regions indicated by thin current layers.
In particular, the presented $z = 0$ surface features two pairs of current layers (located at $[x,y] \sim [300,120]\rho_0,[380,340]\rho_0,[750,560]\rho_0,[830,790]\rho_0$) that in this particular projection appear as asymmetric in-plane magnetic X-points centered on one of their ends, with a large magnetic O-point attached to the other end.
As the magnetic domains (patches of positive and negative out-of-plane $B_z$ field) shift due to the global bulk motions, by $ct/L = 1.54$ (the moment of minimum total magnetic energy) they appear to form diagonal bands.
The pairs of reconnection regions feature non-ideal electric fields, seen as regions of $\bm{E}\cdot\bm{B} \sim -0.2B_0^2$, they are also the main sites of particle heating.
They appear to be connected by common magnetic flux tubes, consequently they approach closer to each other, and eventually they merge around $ct/L = 1.69$ at locations $[x,y] \sim [330,220]\rho_0,[790,680]\rho_0$, boosting the particle mean energy to the levels $\left<\gamma\right> > 15\Theta$.
These localized structures of enhanced current density, particle temperature and $\bm{E}\cdot\bm{B}$ largely disappear by $ct/L = 1.84$ (the moment of peak total electric energy).
The overall effect of magnetic reconnection in the presented 2D section is a gradual growth of magnetic flux wrapped around the 4 largest domains of the out-of-plane magnetic field, as compared with the initial configuration consisting of 8 equally large domains.

Let us now consider this process in more detail, using the $y$ profiles of plasma parameters extracted from the Patch A (Figure \ref{fig_prof1d}).
At $ct/L = 1$, Patch A features a doubly inverted magnetic field component $B_x$ and a singly inverted magnetic field component $B_z$.
By $ct/L = 1.24$, two current layers form at $y/\rho_0 \sim 570,780$. These currents consist of two main components: positive $j_y$ and opposite $j_z$, both peaking at the saturation level of $\pm 0.5ecn_0$.
We should stress here that these current layers are aligned approximately, but not strictly, with the $x$ coordinate, hence $j_y$ is dominated by, but not equal to, the parallel component of the in-plane current density. 
By $ct/L = 1.24$, particles in the middle of the layers are heated to mean energy of $\left<\gamma\right> \simeq 7.5\Theta$, even though the non-ideal field scalar $\bm{E}\cdot\bm{B}$ is still consistent with zero.
The profile of enhanced mean energy $\left<\gamma\right>$ appears to be broader than the profile of enhanced number density $n$, which in turn is similar to the profiles of current density $j_y,j_z$.
The charge density at this stage is at the level of $\rho_e < 0.2 e n_0$.
By $ct/L = 1.39$, the $\bm{E}\cdot\bm{B}$ becomes non-zero, reaching the values of $-0.07 B_0^2$.
By $ct/L = 1.54$, the current layers have shifted towards each other, at the same time they rotate away from the $x$ axis, so that their $y$ profiles appear to be broader.
At this stage, the two layers are separated by a low-density cold region filled mostly with the uniform magnetic field component $B_y$ that connects the two layers, hence there is a pressure cavity and no obstacle that would prevent the two layers from merging.
The merger is observed at $ct/L = 1.69$, and is accompanied by very strong current $j_y \sim 1.5 ecn_0$, rapid variation of the non-ideal field scalar $\bm{E}\cdot\bm{B} \simeq -0.15 B_0^2$, cancellation of magnetic field gradients ($\partial_yB_x,\partial_yB_y$), particle density compression to $n \simeq 2n_0$, and particle heating to $\left<\gamma\right> \simeq 17\Theta$.
By $ct/L = 1.84$, the current merger area becomes very quiet, the current layers, density and temperature structures largely disappear, the magnetic field is dominated by uniform $B_x \simeq B_0$ component and a gradient of the $B_z$ component of opposite sign as compared with the initial state.
The most significant evidence of the recent violent collapse is significant charge imbalance with $\rho_e \sim 0.5 en_0$.

\subsection{Volume distribution}
\label{sec_res_voldist}

\begin{figure*}
\includegraphics[width=0.32\textwidth]{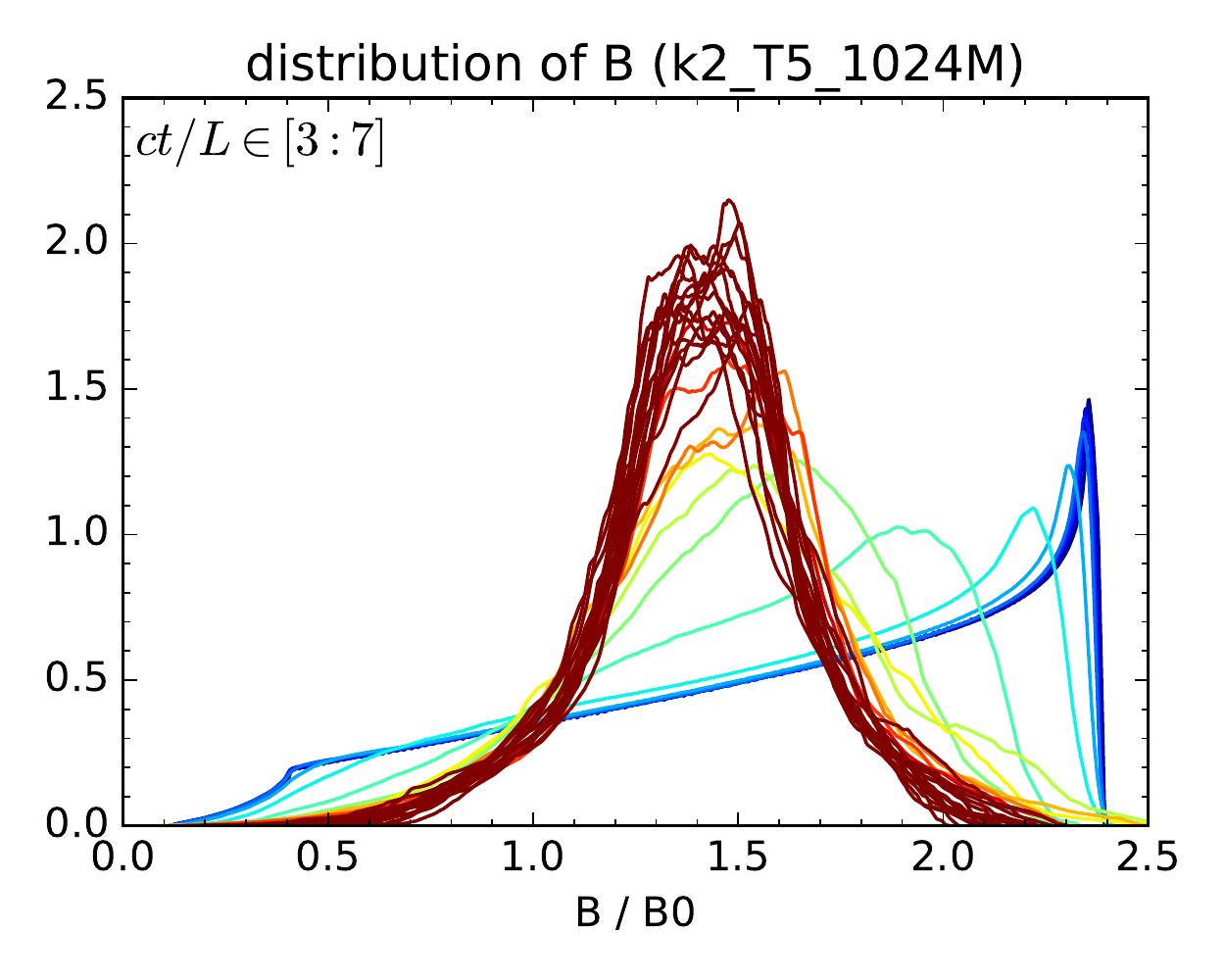}
\includegraphics[width=0.32\textwidth]{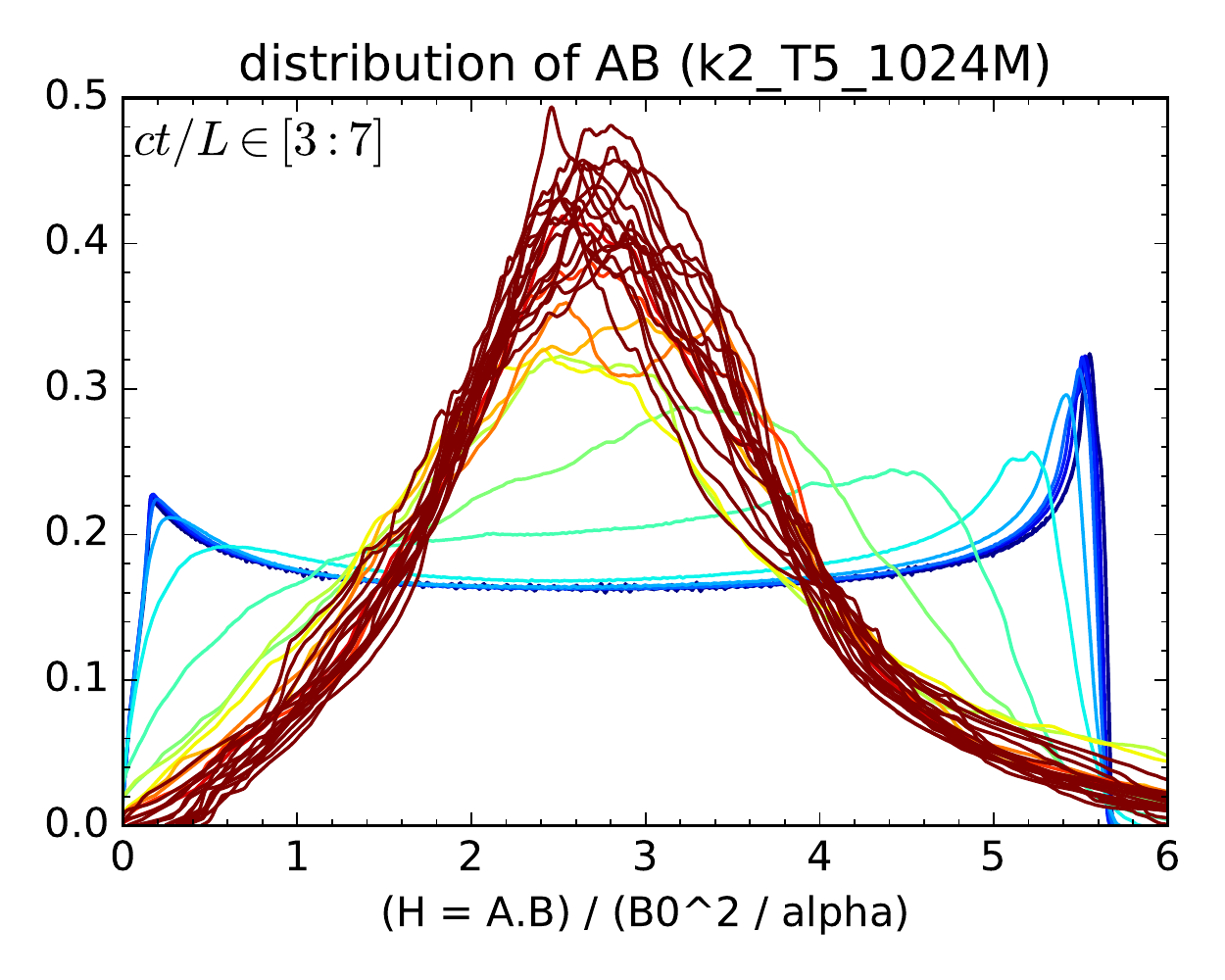}
\includegraphics[width=0.32\textwidth]{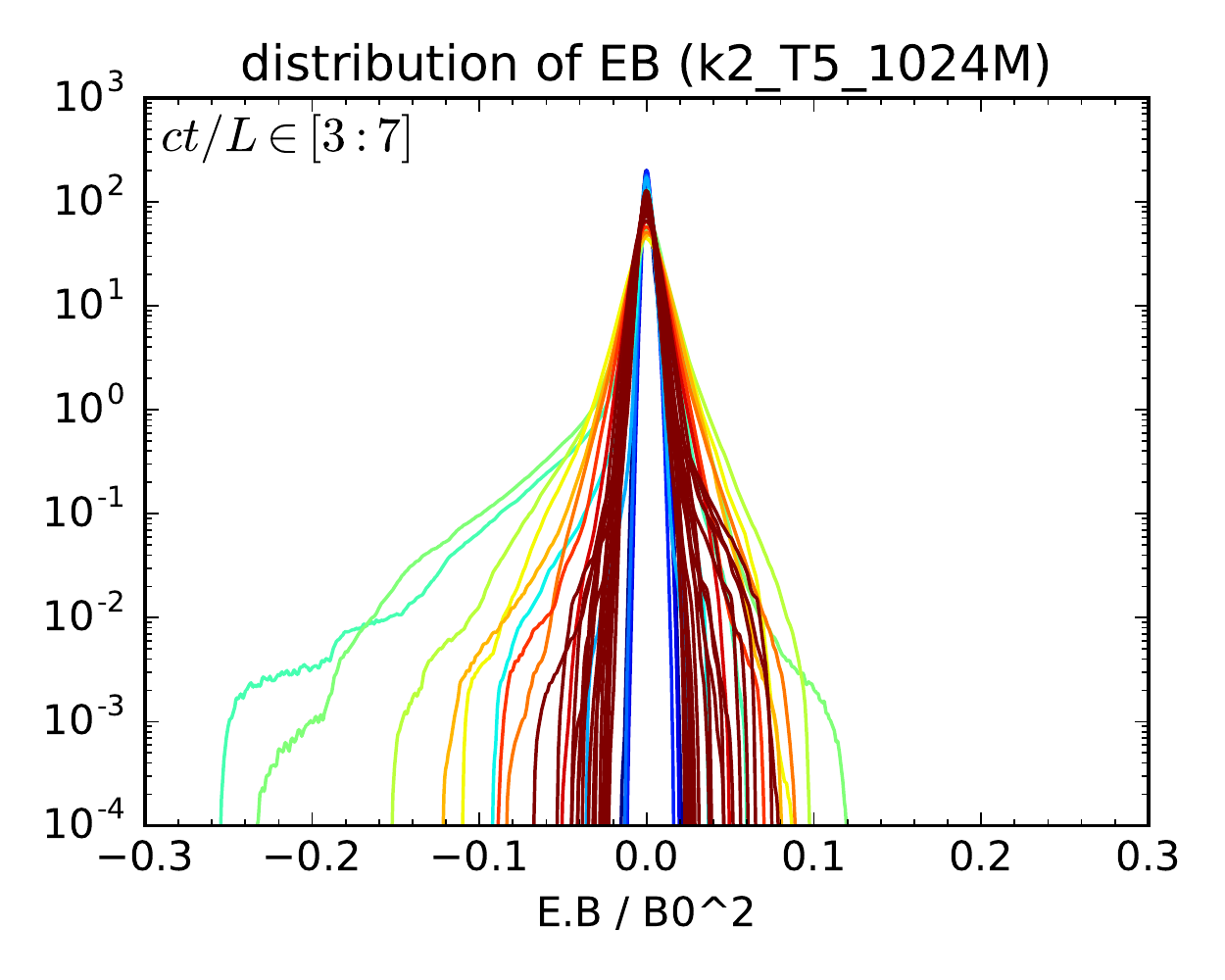}
\\
\includegraphics[width=0.32\textwidth]{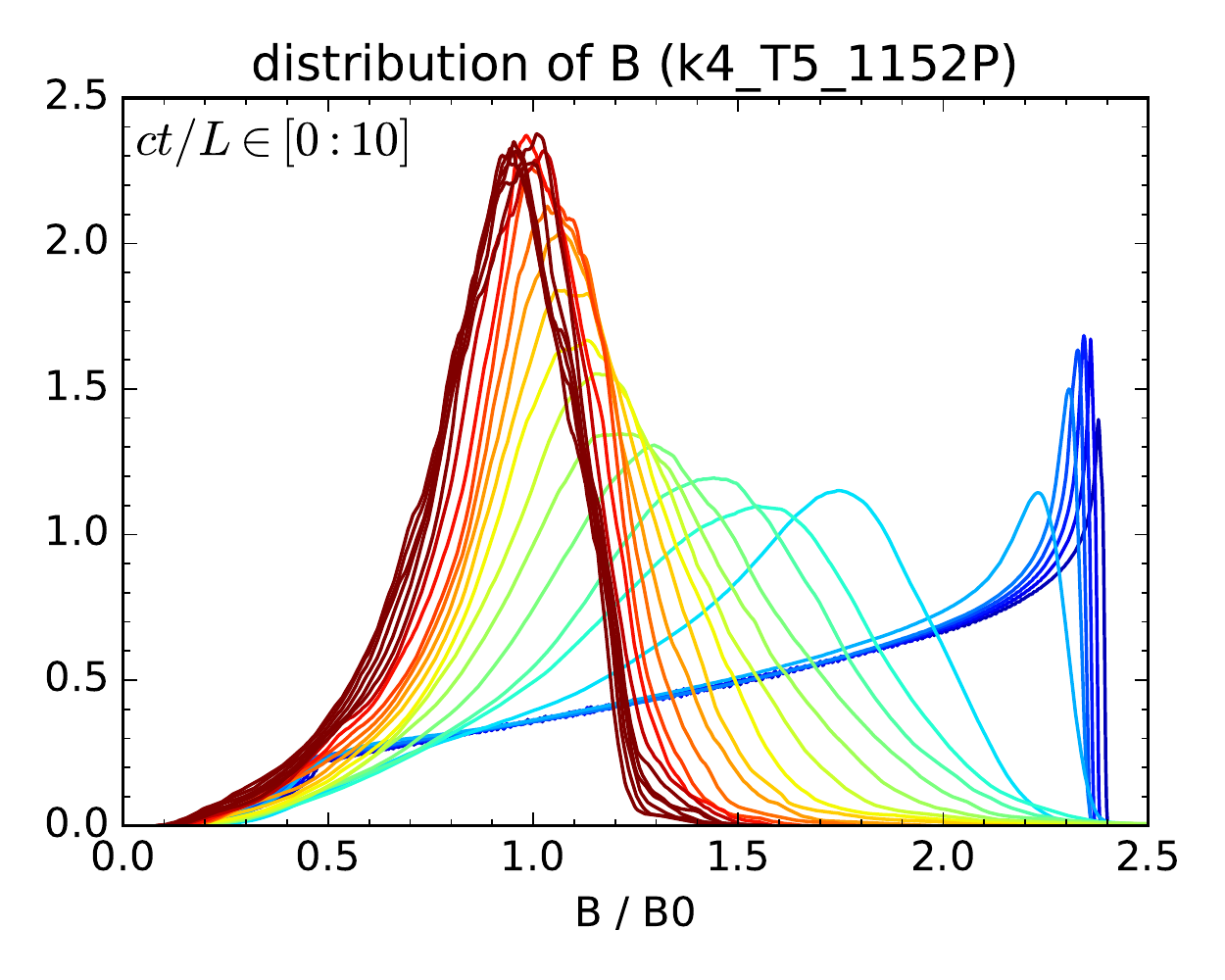}
\includegraphics[width=0.32\textwidth]{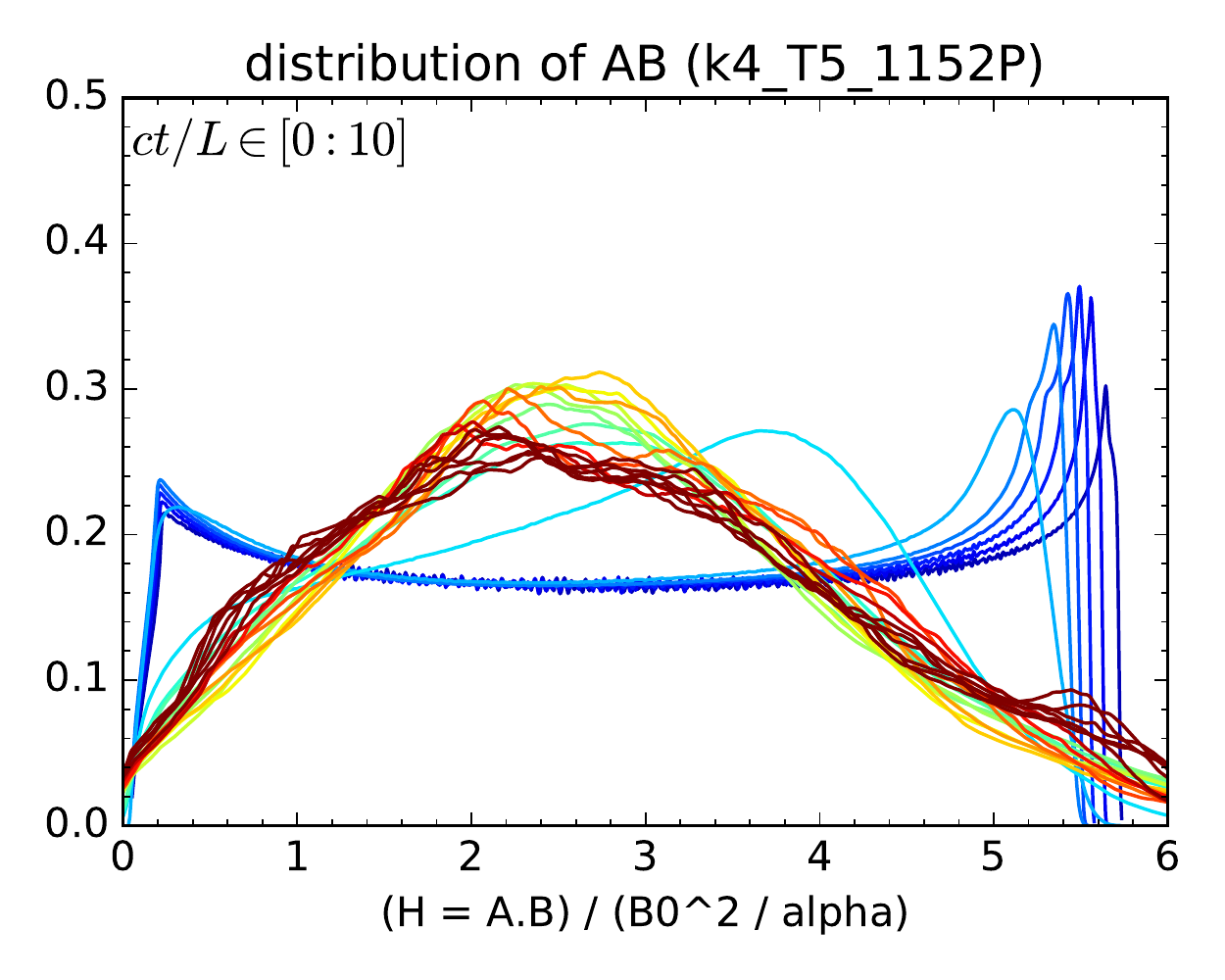}
\includegraphics[width=0.32\textwidth]{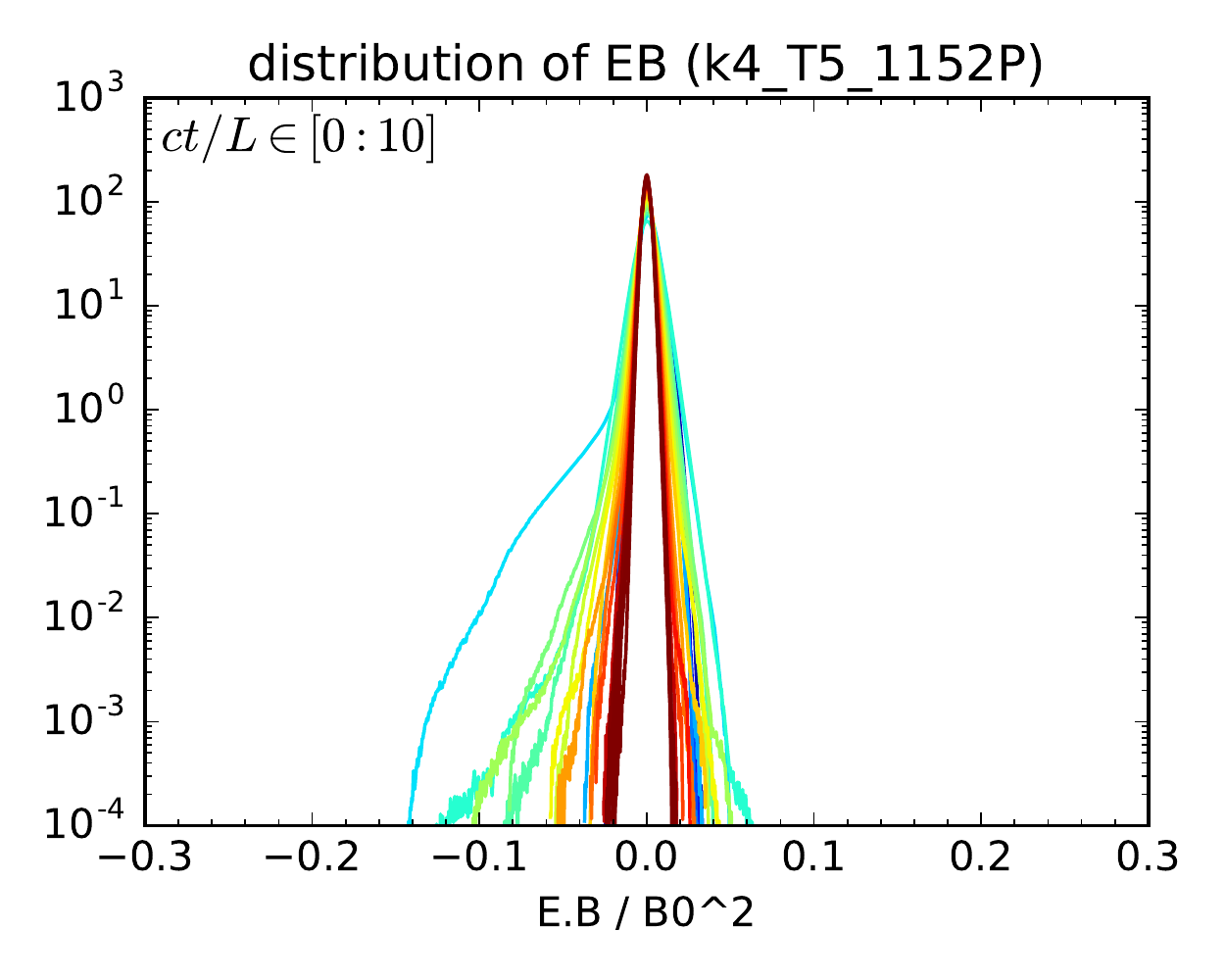}
\caption{Volume probability distributions of magnetic field strength $|\bm{B}| / B_0$ (left panels), magnetic helicity $H / (B_0^2/\alpha_k)$ (middle panels), and non-ideal electric field scalar $(\bm{E}\cdot\bm{B}) / B_0^2$ (right panels) for simulations k2\_T5\_1024M (top panels) and k4\_T5\_1152P (bottom panels). Line colors indicate the progression of simulation time from deep blue to brown within the time ranges indicated in the top left corner of each panel.}
\label{fig_voldist}
\end{figure*}

Figure \ref{fig_voldist} shows volume probability distributions of the magnetic field strength $|\bm{B}|$, magnetic helicity $H = \bm{A}\cdot\bm{B}$, and of the non-ideal field scalar $\bm{E}\cdot\bm{B}$ for simulations k2\_T5\_1024M and k4\_T5\_1152P.
The initial distribution of magnetic field strength is determined by the adopted ABC field configuration, and is skewed towards high values, peaking close to the maximum value $B_{\rm max} = \sqrt{6}B_0$.
During the course of simulation, in the case of $k_{\rm ini} = 2$, the distribution of $|\bm{B}|$ approaches a symmetric distribution with the mean value of $\simeq 1.37 B_0$ and the standard deviation of $\simeq 0.25 B_0$.
However, in the case of $k_{\rm ini} = 4$, the final distribution of $|\bm{B}|$ is asymmetric with the mean value of $\simeq 0.87 B_0$.

For comparison, we show the volume distributions of magnetic helicity, a quantity that is globally conserved in our simulations (see Section \ref{sec_res_totene}). The initial distribution of $H$ is more uniform than that of $B$, extending up to $H_{\rm max} \simeq 5.7 B_0^2/\alpha_k$.
While the mean value of $\left<H\right> \simeq 3B_0^2/\alpha_k$ is roughly conserved (see Figure \ref{fig_tot_ene}), the final distribution of $H$ is narrower and approximately symmetric, with the standard deviation of $\simeq 1 B_0^2/\alpha_k$ for simulation k2\_T5\_1024M, and $\simeq 1.45 B_0^2/\alpha_k$ for simulation k4\_T5\_1152P. The relaxed shape of the final distribution of $H$ suggests that the final state is not consistent with the ABC configuration for $k = 1$.

The distributions of $\bm{E}\cdot\bm{B}$ are strongly concentrated at the zero value. They were calculated after Gaussian smoothing of the $\bm{E}\cdot\bm{B}$ volume data cubes with radius of 3 cells in order to suppress the Poisson noise in the electric field. A temporary excess of negative values, up to $\bm{E}\cdot\bm{B} \simeq -0.25B_0^2$ in the case of $k_{\rm ini} = 2$ and up to $\bm{E}\cdot\bm{B} \simeq -0.14B_0^2$ in the case of $k_{\rm ini} = 4$, is seen for simulation times corresponding to the most rapid dissipation of magnetic energy.

We also performed a Fourier decomposition of magnetic field distribution with two main goals: (1) to evaluate the contribution of individual dominant modes to the magnetic energy, (2) to characterize the turbulent cascade of magnetic energy. The magnetic energy spectrum is calculated as $\mathcal{E}_{B,k} = |\hat{B}_k|^2$, where $\hat{B}_k$ is the Discrete Fourier Transform (DFT) of the magnetic field strength $B(x,y,z)$ for $k \in \{0,1,...,N_x/2-1\}(2\pi/L)$, normalized to the total magnetic energy $\sum_k\mathcal{E}_{B,k} = \sum_{x,y,z}B^2(x,y,z)$.

\begin{figure*}
\includegraphics[width=0.32\textwidth]{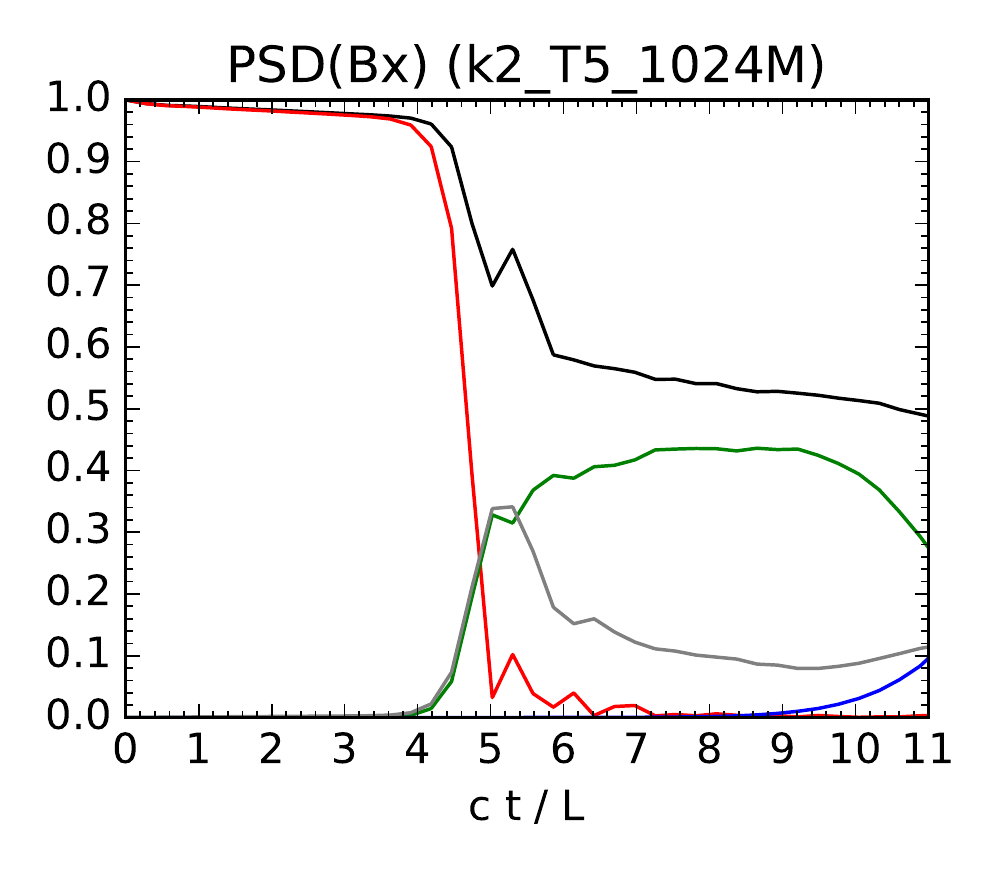}
\includegraphics[width=0.32\textwidth]{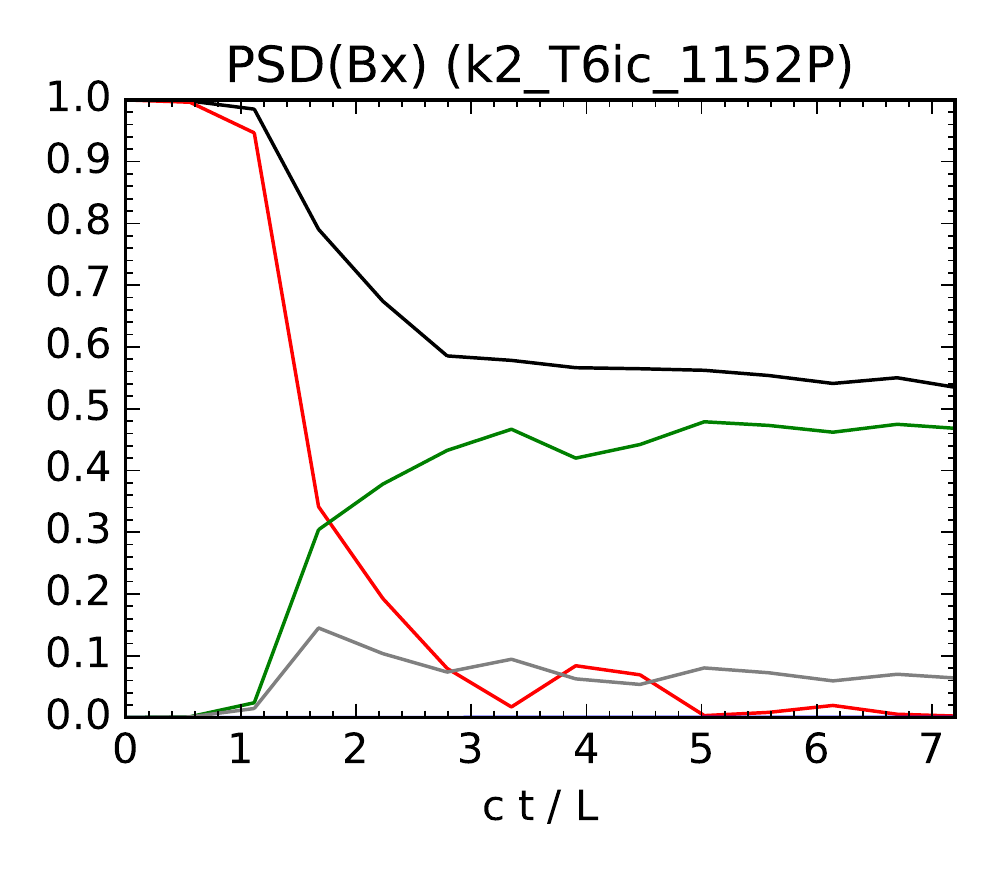}
\includegraphics[width=0.32\textwidth]{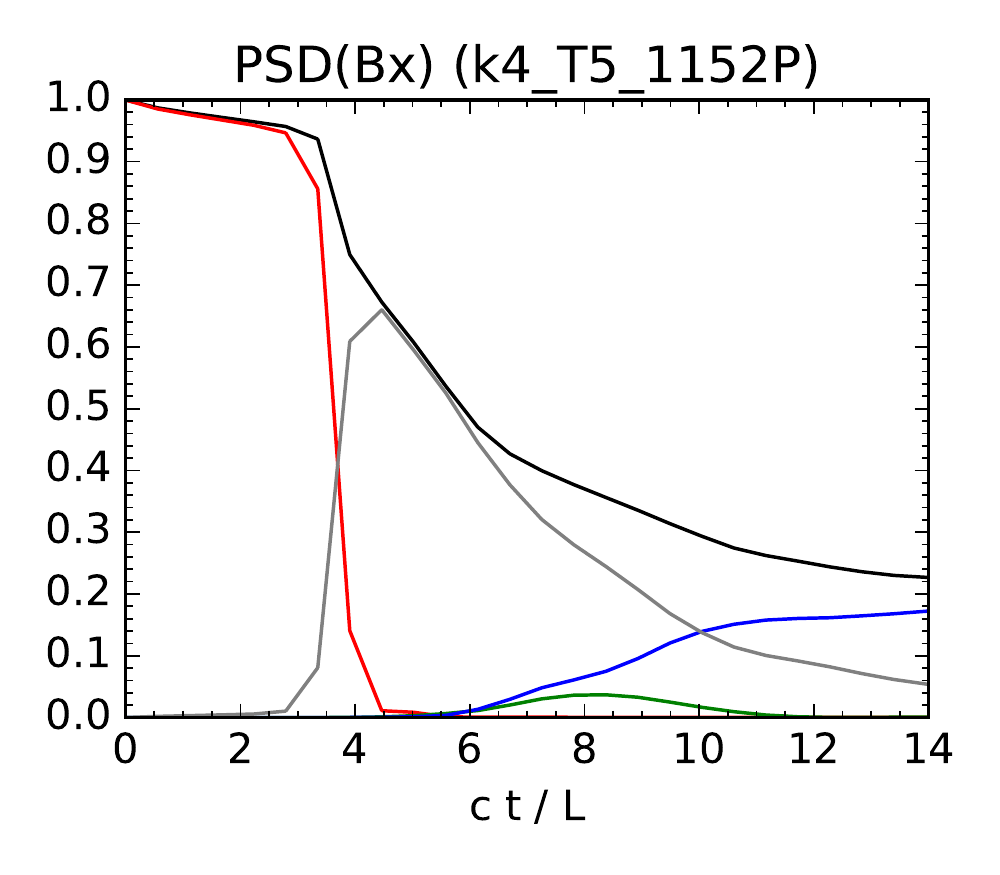}
\caption{Energy contribution of the dominant Fourier modes of the $B_x$ magnetic field component as function of simulation time $ct/L$ for simulations k2\_T5\_1024M (left), k2\_T6ic\_1152P (middle), and k4\_T5\_1152P (right). The black lines show the total energy of the $B_x$ component, represented as $\left<B_x^2\right>$. The red lines show the contribution of the initial ``ABC'' modes: $(0,k_{\rm ini},0)$ and $(0,0,k_{\rm ini})$; the green lines show the contribution of the $(1,1,0)$ modes; the blue lines show the contribution of the ``Taylor'' modes $(0,1,0)$ and $(0,0,1)$; and the gray lines show the residual energy contained in all other modes.}
\label{fig_psd_modes}
\end{figure*}

Figure \ref{fig_psd_modes} shows the energy contributions of the dominant modes obtained by Fourier decomposition of the volume distribution of the $B_x$ component (decomposition of other components $B_y$ and $B_z$ yields equivalent results, as our simulations are isotropic, while decomposition of the magnetic field strength $|\bm{B}|$ is dominated by the uniform mode $(k_x=0,k_y=0,k_z=0)$) as functions of simulation time for simulations k2\_T5\_1024M, k2\_T6ic\_1152P and k4\_T5\_1152P (these are our longest simulations running up to $ct/L \sim 7-14$). The initial ABC modes $(0,k_{\rm ini},0)$ and $(0,0,k_{\rm ini})$ are rapidly and completely destroyed during the main energy dissipation phase. For $k_{\rm ini} = 2$, the most important emerging mode is $(1,1,0)$, although in the simulation k2\_T5\_1024M we observe the $k=1$ ABC modes $(0,1,0)$ and $(0,0,1)$ emerging for $ct/L > 9$. For $k_{\rm ini} = 4$, the $k=1$ ABC modes dominate the final state, while the transition features numerous second-order modes $(0,2,2)$, $(2,0,2)$, $(1,0,2)$, $(0,1,2)$, etc., while the $(1,1,0)$ mode is not important. We should therefore conclude that magnetic relaxation towards the $k=1$ ABC state is more advanced in our $k_{\rm ini} = 4$ simulation.

\begin{figure*}
\includegraphics[width=0.32\textwidth]{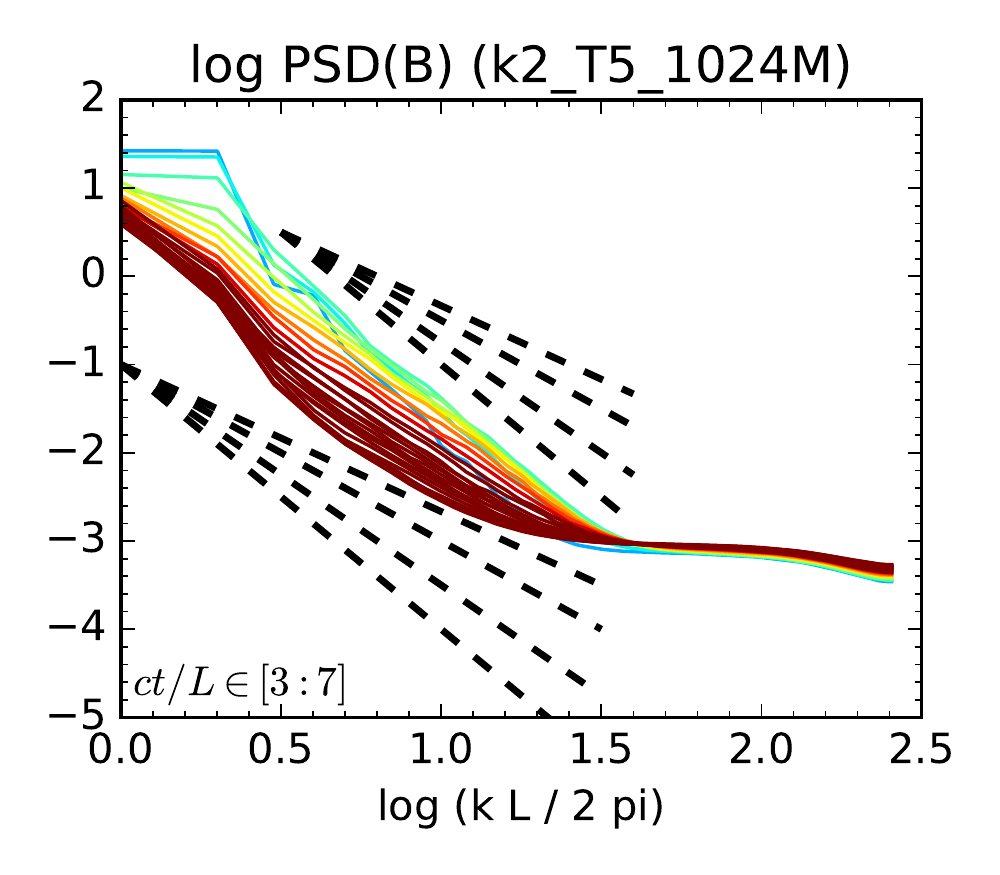}
\includegraphics[width=0.32\textwidth]{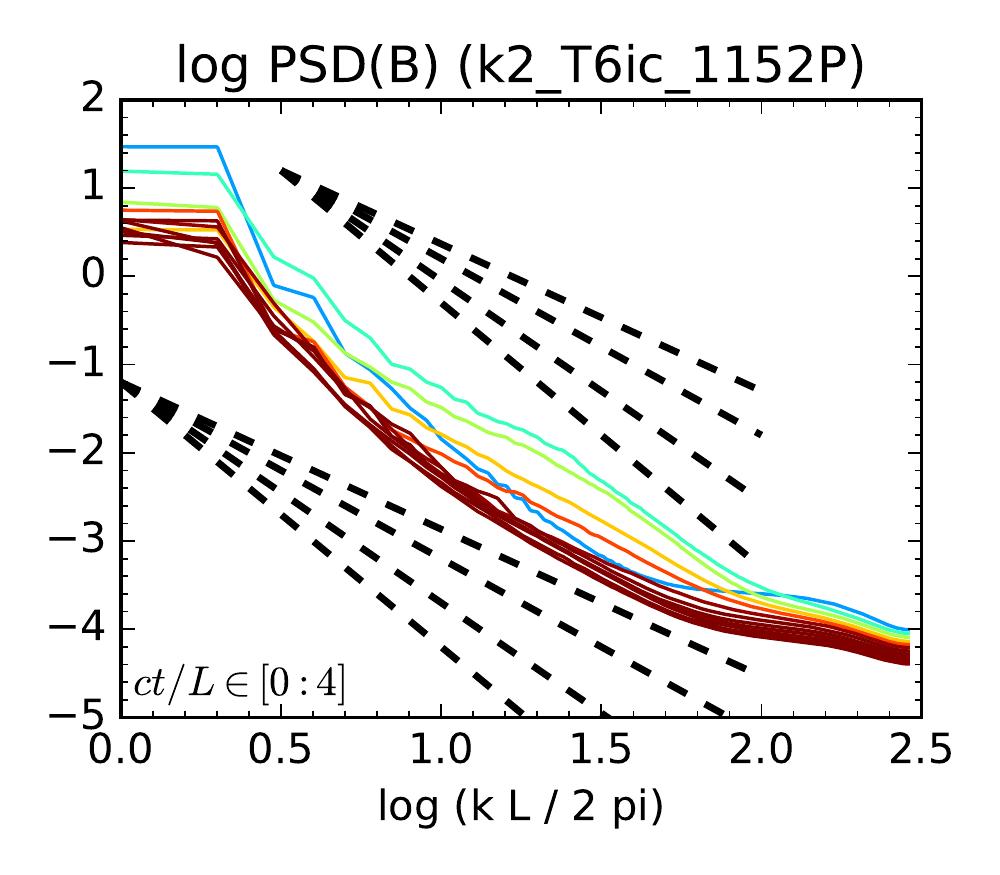}
\includegraphics[width=0.32\textwidth]{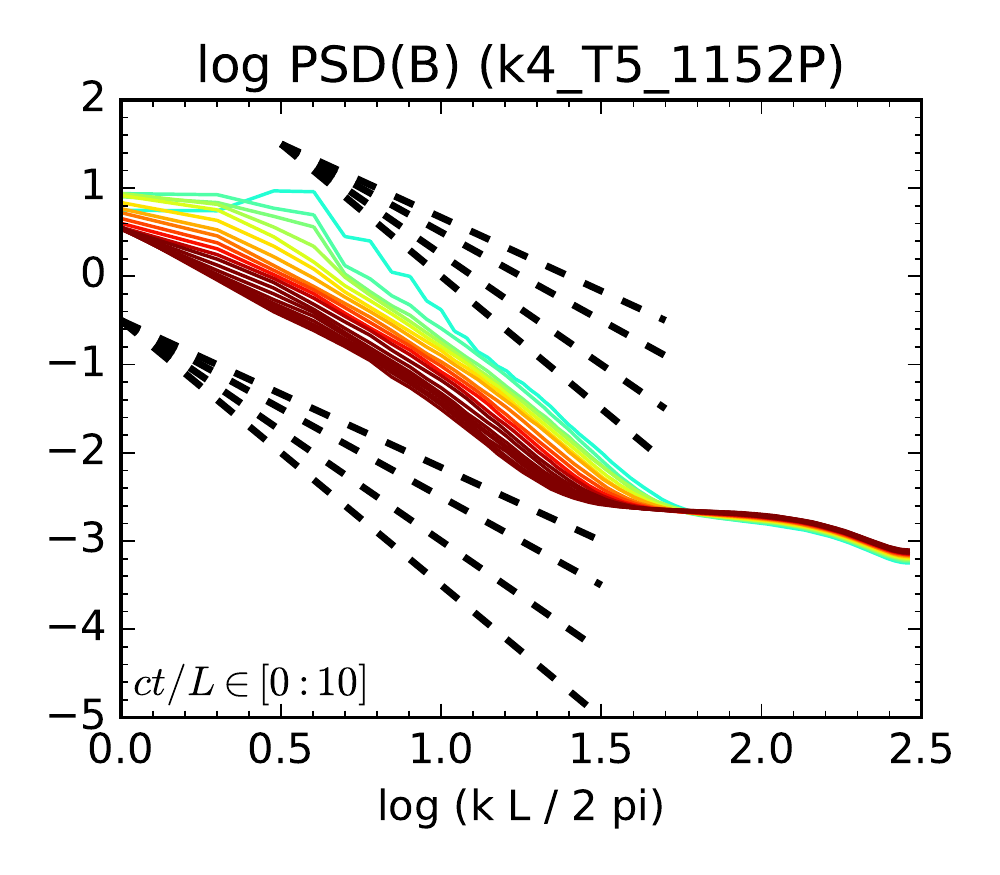}
\caption{Magnetic energy spectra $\mathcal{E}_{B,k} = |\hat{B}_k|^2$ calculated for simulations k2\_T5\_1024M (left), k2\_T6ic\_1152P (middle), and k4\_T5\_1152P (right). Line colors indicate the progression of simulation time from deep blue to brown within the time ranges indicated in the bottom left corner of each panel. The thick dashed lines indicate the power-law slopes of $-5/3,-2,-2.5,-3$.}
\label{fig_psd}
\end{figure*}

Figure \ref{fig_psd} shows the magnetic energy spectra for simulations k2\_T5\_1024M, k2\_T6ic\_1152P and k4\_T5\_1152P for a range of simulation times.
The magnetic energy spectra are dominated by the white noise for $k > 40(2\pi/L)$ for the T5 simulations and for $k > 100(2\pi/L)$ for the k2\_T6ic\_1152P simulation.
The inertial subrange extends over at least one order of magnitude in wavenumber $3 < kL/(2\pi) < 30(100)$.
In all cases, we observe a freely decaying turbulent cascade that can be described roughly as a power-law $\mathcal{E}_{\rm B}(k) \propto k^{-2.5}$, with significant deviations in either direction.
The cascade appears to be most regular for simulation k4\_T5\_1152P.

Fourier decomposition was also attempted for electric field $\bm{E}$, and for velocity field $\bm{v}$, however, the corresponding spectra of electric and kinetic energies are dominated by the white noise.

\subsection{Particle momentum distribution}
\label{sec_res_partacc}

\begin{figure*}
\includegraphics[width=0.32\textwidth]{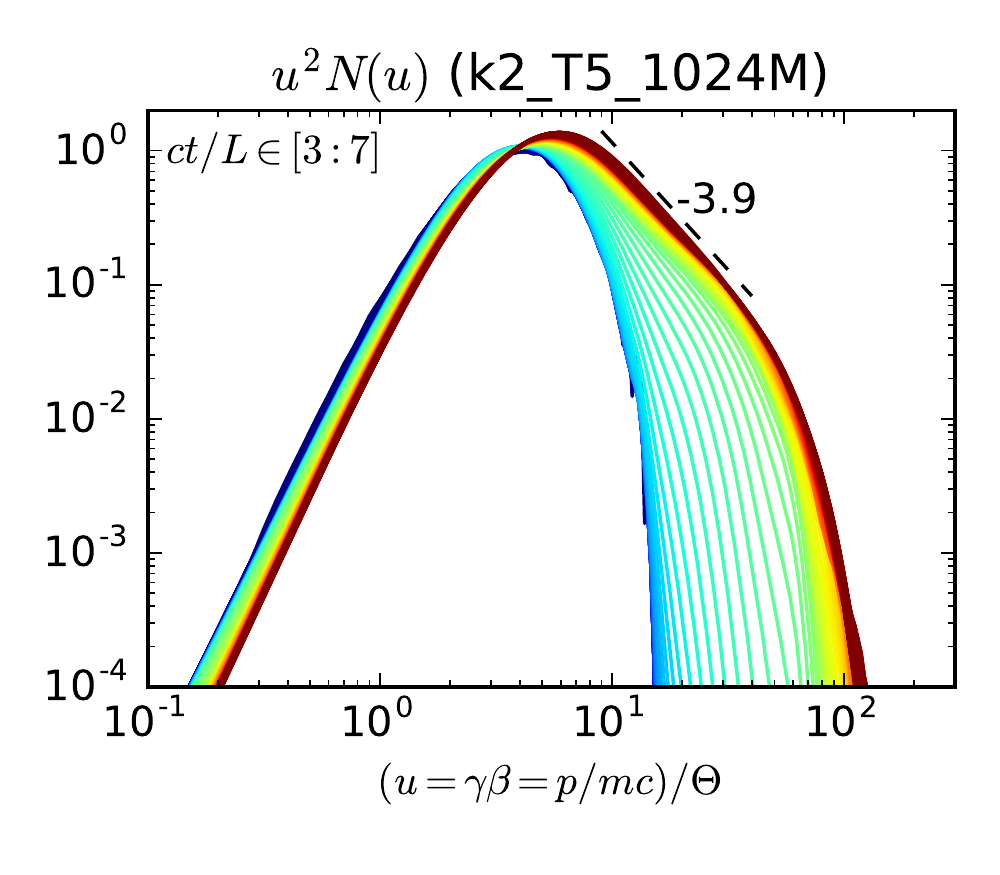}
\includegraphics[width=0.32\textwidth]{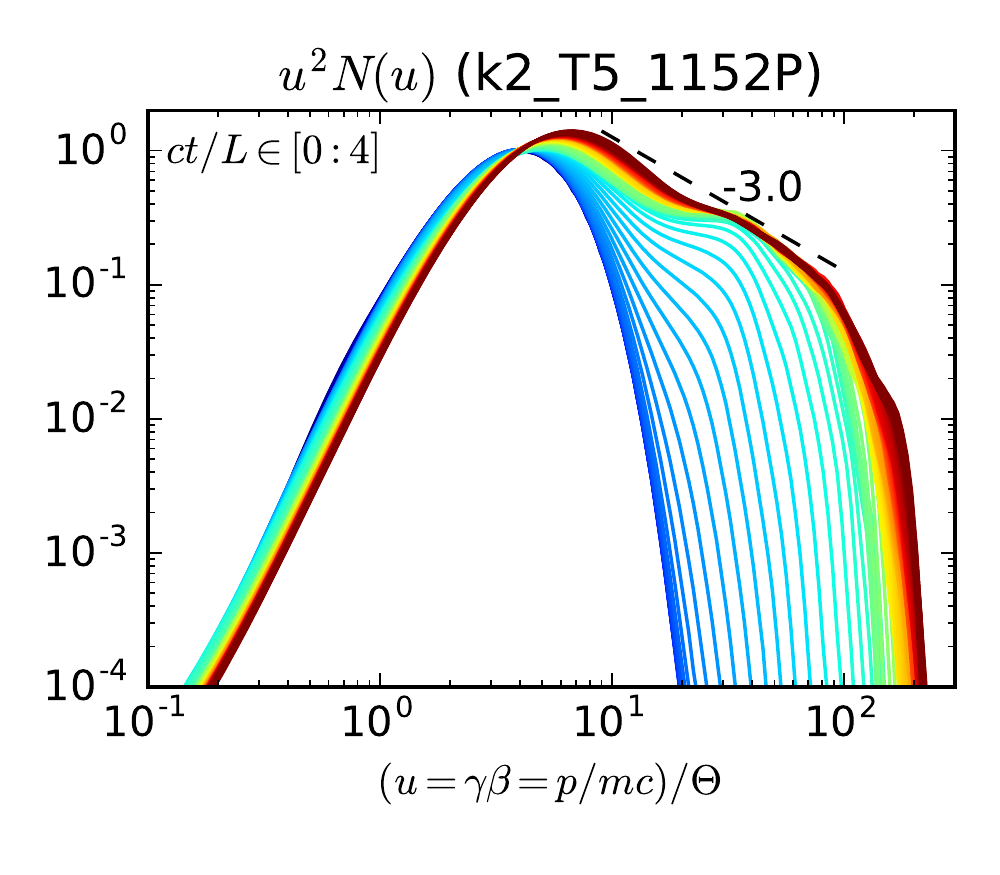}
\includegraphics[width=0.32\textwidth]{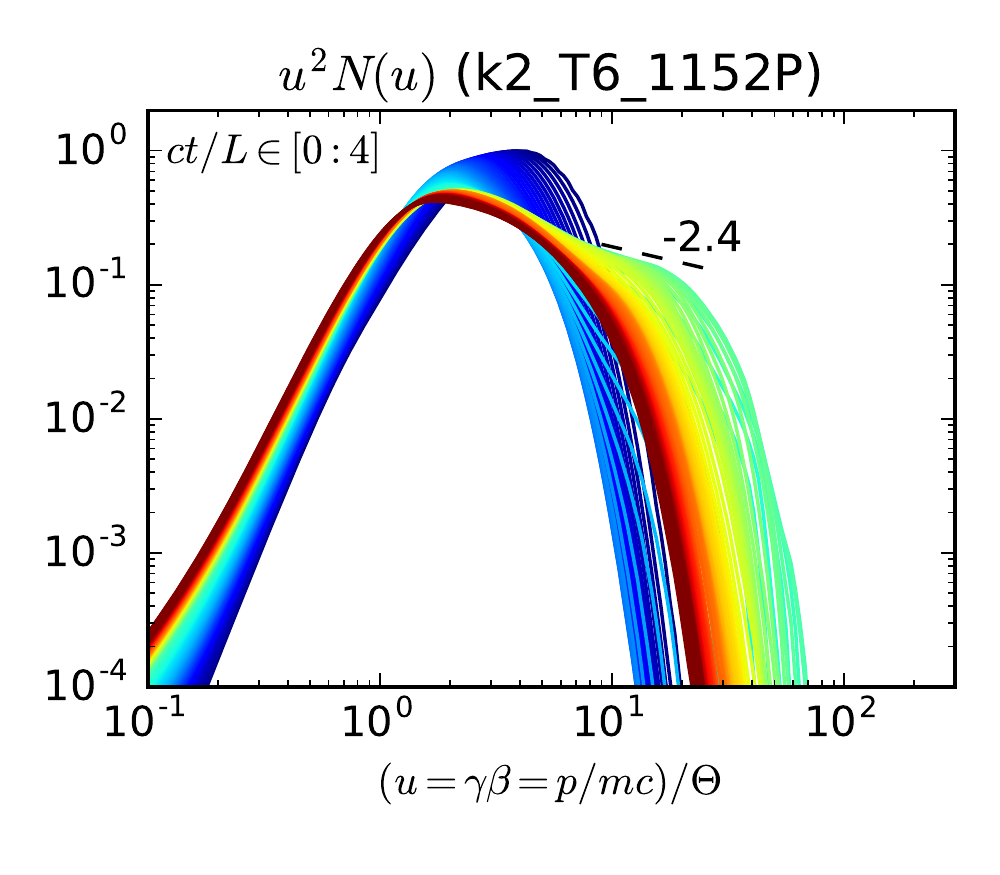}
\includegraphics[width=0.32\textwidth]{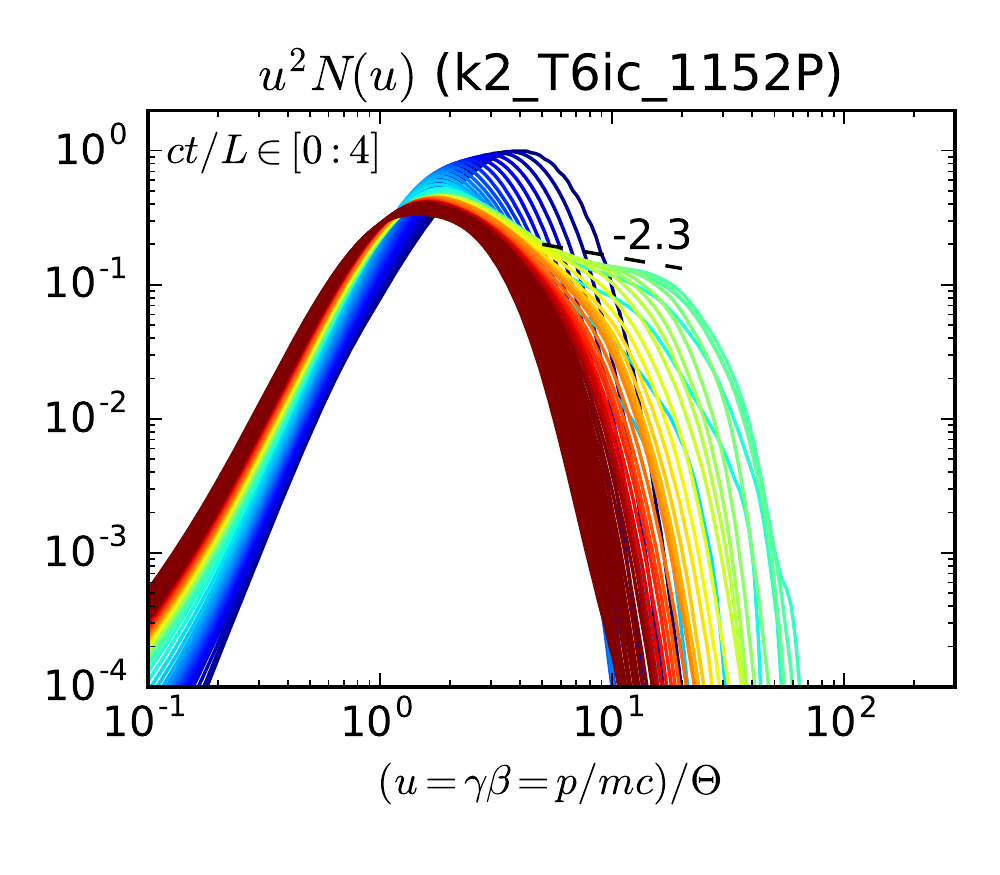}
\includegraphics[width=0.32\textwidth]{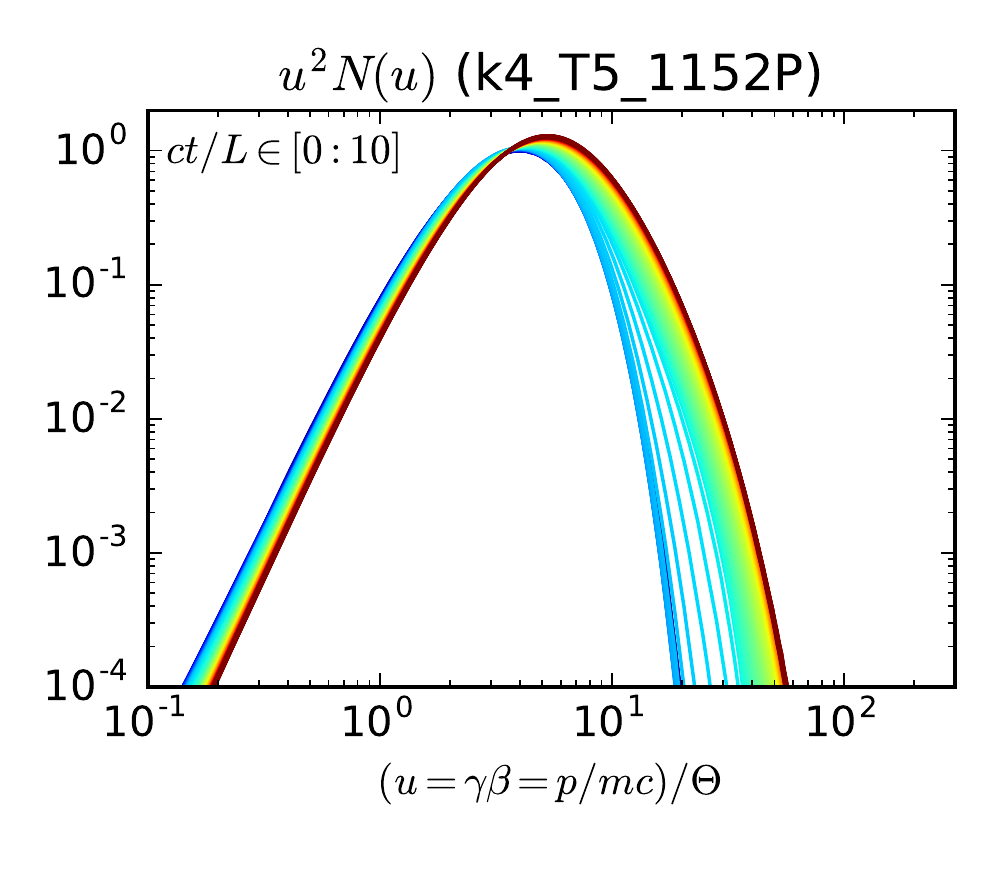}
\caption{Particle momentum distributions $u^2 N(u)$ compared for all simulations. The distributions are normalized to the peak of the initial distribution for each simulation. Line colors indicate the progression of simulation time from deep blue to brown over the range indicated in the top left corner of each panel.}
\label{fig_part_spe}
\end{figure*}

Figure \ref{fig_part_spe} presents the time evolution of the particle momentum distributions for all 5 simulations. For ultra-relativistic electrons with $\Theta \gg 1$, the dimensionless particle momentum is basically equivalent to the dimensionless particle energy $u = \gamma\beta = p/(m_{\rm e}c) \simeq \gamma$. For all simulations, the initial momentum distribution is adopted to be the Maxwell-J\"{u}ttner distribution (Section 2).

We find evidence for non-thermal particle acceleration for all simulations with $k_{\rm ini} = 2$.
The effective power-law indices $p$ (such that $N(u) \propto u^{-p}$) are estimated roughly by finding a power-law slope that makes a line parallel to the data (see \citealt{Wer16} for more elaborate fitting methods), with a different approach used depending on the radiative cooling efficiency.
In the slow-cooling cases ($\Theta = 10^5$), we measure the shape of the final particle distribution, and we find $p \simeq 3.9$ for simulation k2\_T5\_1024M and $p \simeq 3.0$ for simulation k2\_T5\_1152P.
In the fast-cooling cases ($\Theta = 10^6$), the high-energy excess is short-lived, and appears to be a distinct spectral component, harder than the line joining it with the peak of the low-energy component.
We measure the slope of the high-energy excess at the moment of its most prominent extension, finding $p \simeq 2.4$ for simulation k2\_T6\_1152P and $p \simeq 2.3$ for simulation k4\_T6ic\_1152P.

Simulation k4\_T5\_1152P shows a mild increase of the mean particle energy and a spectral broadening exceeding the shape of the Maxwell-J\"{u}ttner distribution. However, no power-law component can be clearly seen in the high-energy end of the distribution. This simulation is characterized by the lowest value of mean hot magnetization $\left<\sigma_{\rm hot}\right> = 0.6$, which explains inefficient particle acceleration.

\begin{figure*}
\includegraphics[width=\textwidth]{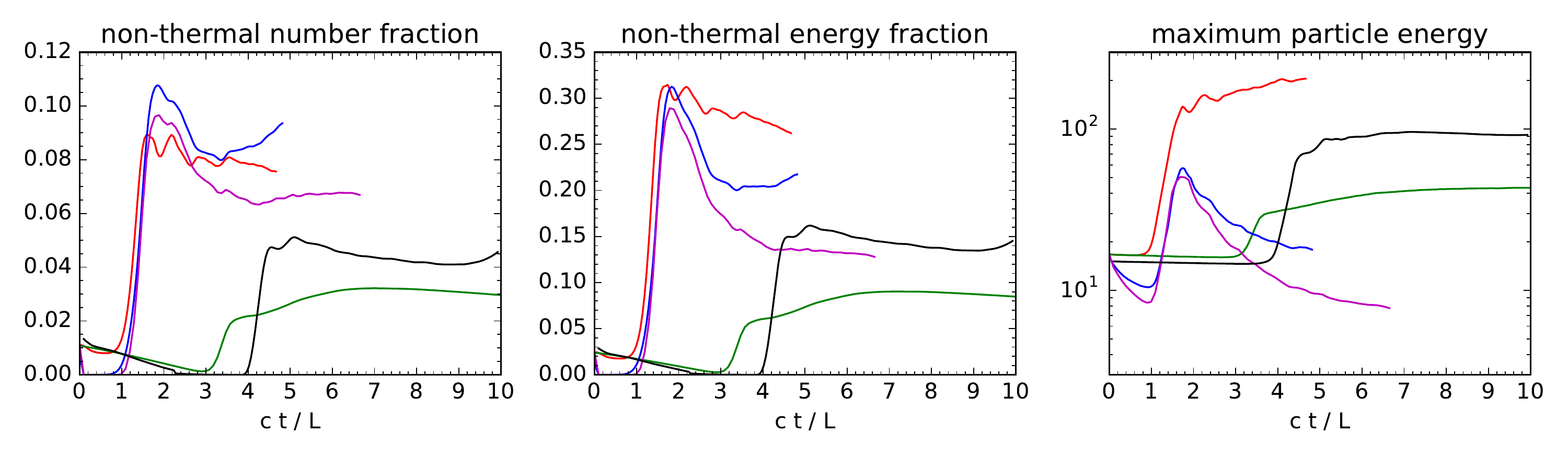}
\caption{Fractions of the particle number $f_n$ (left panel) and energy $f_e$ (middle panel) contained in the non-Maxwellian high-energy distribution tail, and the maximum particle energy $\gamma_{\rm max}/\Theta$ (right panel) evaluated at the $10^{-3}$ level of the normalized $u^2 N(u)$ distribution, as functions of simulation time $ct/L$ compared for all simulations. The line styles are the same as in Figure \ref{fig_tot_ene}.}
\label{fig_part_frac}
\end{figure*}

Figure \ref{fig_part_frac} shows the fractions of particle number $f_n$ and energy $f_e$ contained in the non-thermal high-energy tails of their momentum distributions as function of simulation time.
These fractions are obtained by fitting a Maxwell-J\"{u}ttner distribution to the actual distributions using weights proportional to $u^{-2}$, and subtracting both the integrated contribution of the fitted model, as well as any low-energy excess.
For the $k_{\rm ini} = 2$ simulations k2\_T5\_1152P, k2\_T6\_1152P and k2\_T6ic\_1152P, the non-thermal number fraction reaches $f_n \sim (9-11)\%$, and the non-thermal energy fraction reaches $f_e \sim (28-32)\%$ by $ct/L \simeq 2$. At later times, the energy fraction decays significantly in the fast-cooling cases ($\Theta = 10^6$), approaching the level of $f_e \simeq 13\%$ for simulation k2\_T6ic\_1152P. For simulation k2\_T5\_1024M, the fractions reach $f_n \simeq 5\%$ and $f_e \simeq 16\%$ by $ct/L \simeq 5$. And for simulation k4\_T5\_1152P, the fractions only reach $f_n \simeq 3\%$ and $f_e \simeq 9\%$ at $ct/L \sim 7$.

Figure \ref{fig_part_frac} also shows the maximum particle energy $\gamma_{\rm max}$, normalized to the initial particle dimensionless temperature $\Theta$ and evaluated at the level of $10^{-3}$ of the $u^2 N(u)$ distribution with its peak normalized to unity at $t = 0$, as presented in Figure \ref{fig_part_spe}.
For the cases with lower initial particle temperature $\Theta = 10^5$, the fact that maximum particle energy does not decrease confirms that radiative cooling is not efficient.
We find the following values: $\gamma_{\rm max} \simeq 200$ for simulation k2\_T5\_1152P, $\gamma_{\rm max} \simeq 90$ for simulation k2\_T5\_1024M, and $\gamma_{\rm max} \simeq 40$ for simulation k4\_T5\_1152P.
For the cases with higher initial particle temperature $\Theta = 10^6$, we find significant decrease of $\gamma_{\rm max}$ in time, both from the peak values of $\gamma_{\rm max} \simeq 55$, and from the initial value of $\simeq 16$. For simulation k2\_T6ic\_1152P, the final obtained value is $\gamma_{\rm max} \simeq 8$.

\subsection{Particle angular distribution}
\label{sec_res_partang}

Particle acceleration by relativistic magnetic reconnection is characterized by strong energy-dependent particle anisotropy, such effect is termed \emph{kinetic beaming} \citep{Cer12}, as opposed to the Doppler beaming that effects equally particles of all energies.
Kinetic beaming has also been demonstrated in 2D simulations of ABC fields by \cite{Yua16}, and we now assert that it is also present in our 3D simulations.
Figure \ref{fig_tot_maps} shows the angular distributions of the most energetic electrons and positrons, selected for their momentum values $u > 40\Theta$, at several moments of simulation k2\_T5\_1152P.
We find two broad and narrow beams/fans at $ct/L = 1.54,1.69$ located symmetrically and away from the cardinal directions $\pm\hat{\bm x},\pm\hat{\bm y},\pm\hat{\bm z}$.
One of them stretches from the $xz$ plane between $+\hat{\bm x}$ and $-\hat{\bm z}$ to the $yz$ plane between $+\hat{\bm y}$ and $+\hat{\bm z}$.
The two fans cross the $xy$ plane either for positive $x$ and positive $y$, or for negative $x$ and negative $y$, this is consistent with the orientations of current layers in the $z = 0$ surface.
Interestingly, at $ct/L = 1.54$ both fans are fragmented into 4 smaller structures that most likely correspond to different reconnection sites across the simulation volume.
In fact, the angular distribution of energetic particles is subject to rapid and complicated variations (not simply the dilemma of spatial bunching vs. beam sweeping), as can be seen in the supplementary movie \ref{movie_part_ang}, that predict complex lightcurves seen by different observers.

\subsection{Individual energetic particles}
\label{sec_res_partind}

\begin{figure*}
\includegraphics[width=\textwidth]{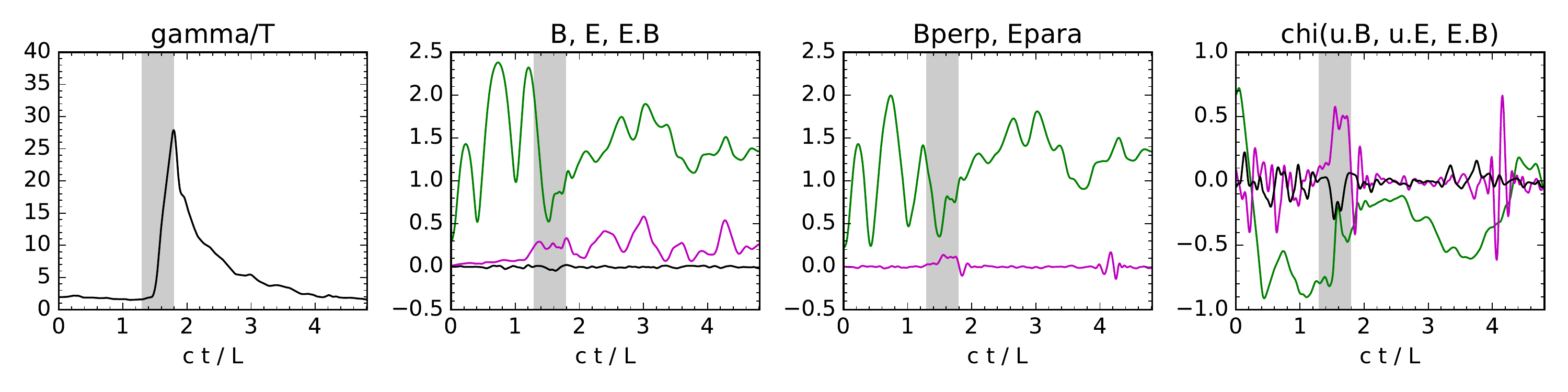}
\caption{Example of acceleration history for an individual tracked high-energy positron, taken from simulation k2\_T6\_1152P with strong synchrotron cooling (see Supplementary Movie \ref{movie_orbits} for more examples). The first panel from the left shows the particle energy $\gamma/\Theta$ as function of $ct/L$. The second panel shows the magnetic field strength $B$ (green line), the electric field strength $E$ (magenta line) and the $\bm{E}\cdot\bm{B}$ scalar (black line), with all the field values interpolated to the instantaneous particle position. The third panel shows the magnetic field component perpendicular to the particle momentum $B_\perp = |\bm{B}\times(\bm{u}/u)|$ (green line), and the electric field component parallel to the particle momentum $E_\parallel = \bm{E}\cdot(\bm{u}/u)$ (magenta line). The last panel shows the cosines of angles between the $\bm{u},\bm{B},\bm{E}$ vectors: $\chi_{u,B} = (\bm{u}/u)\cdot(\bm{B}/B)$ (green line), $\chi_{u,E} = (\bm{u}/u)\cdot(\bm{E}/E)$ (magenta line), and $\chi_{E,B} = (\bm{E}/E)\cdot(\bm{B}/B)$ (black line). The main acceleration episode (MAE) is indicated with the gray rectangles.}
\label{fig_orbit_example}
\end{figure*}

For each simulation, we have tracked around $4\times 10^4$ individual electrons and positrons, recording their positions $\bm{r}(t)$, momenta $\bm{u}(t)$, and the values of magnetic and electric vectors $\bm{B}(t),\bm{E}(t)$ interpolated to $\bm{r}(t)$. Of these, we selected the subsamples of energetic tracked particles that exceeded at any moment the energy threshold of $\gamma_{\rm min} = 20\Theta$. Figure \ref{fig_orbit_example} shows an example of individual tracked positron recorded in simulation k2\_T6\_1152P characterized by efficient synchrotron cooling. This particle reaches a maximum energy of $\gamma_{\rm max} \simeq 28\Theta$ at $ct/L \simeq 1.8$, and it cools down to the initial energy of $\gamma \simeq 2\Theta$ by $ct/L \simeq 4$. Using the particle energy history $\gamma(t)$, after applying a Gaussian smoothing with dispersion $\sigma_{ct/L} \simeq 0.01$, we identify the \emph{main acceleration episode (MAE)} as the contiguous period of time $[t_1:t_2]$ corresponding to the largest monotonic energy gain $\Delta\gamma = \gamma(t_2)-\gamma(t_1)$. The MAE boundary times $ct_1/L \simeq 1.3$ and $ct_2/L \simeq 1.8$ are indicated in Figure \ref{fig_orbit_example}.
This particle was accelerated by electric field inclined at an angle $\arccos[(\bm{u}/u)\cdot(\bm{E}/E)] \simeq 60^\circ$, with the parallel component peaking at $E_\parallel = \bm{E}\cdot(\bm{u}/u) \simeq 0.12B_0$, at insignificant values of $\bm{E}\cdot\bm{B}$ (with $\bm{E}$ and $\bm{B}$ making an angle $\sim 90^\circ - 105^\circ$), and with slightly lower-than-average perpendicular magnetic field $B_\perp = |\bm{B}\times(\bm{u}/u)|$ (with $\bm{B}$ and $\bm{u}$ making an angle $\sim 105^\circ - 120^\circ$).
More examples of individual tracked particles are presented in the Supplementary Movie \ref{movie_orbits}.

Figure \ref{fig_orbits_stat1} shows the distributions of various parameters averaged over the MAE of individual energetic particles.
The distribution of $\Delta\gamma/\gamma_{\rm max}$ is a measure of monotonicity of particle acceleration, i.e. the fraction of total energy gain that can be attributed to the MAE. For simulations with inefficient radiative cooling ($\Theta = 10^5$), the values of $\Delta\gamma/\gamma_{\rm max}$ extend down to $\sim 20\%$, while for simulations with efficient radiative cooling they are typically above $\sim 60\%$. This suggests that the concept of MAE is more useful for acceleration under strong radiative cooling.
Typical values of energy gain during the MAE are $\Delta\gamma \sim (15-20)$ for all simulations.
Time durations of MAE are systematically longer for simulation k2\_T5\_1024M initiated without an explicit magnetic perturbation ($\Delta t \sim 1.5(L/c)$) than for simulations k2\_T*\_1152P ($\Delta t \sim (0.3-0.6)(L/c)$).
The effective acceleration electric fields are the highest for simulation k2\_T5\_1152P ($\left<E_\parallel\right>_t \sim 0.06B_0$) and the lowest for simulation k2\_T5\_1152P ($\left<E_\parallel\right>_t \sim 0.03B_0$).
The component of electric field parallel to the magnetic field is in general lower than the effective acceleration electric field, which means that either the field scalar $\bm{E}\cdot(\bm{B}/B)$ does not capture the entire non-ideal electric field, or that particle acceleration can be partially attributed to the bulk plasma motions, e.g. in colliding pairs of current layers described in Section \ref{sec_res_maps}.
Finally, the typical values of perpendicular magnetic field component are found to be slightly less than $B_0$, down to $\left<B_\perp\right>_t \simeq 0.7B_0$ for simulations k2\_T6*\_1152P.

\begin{figure*}
\includegraphics[width=0.32\textwidth]{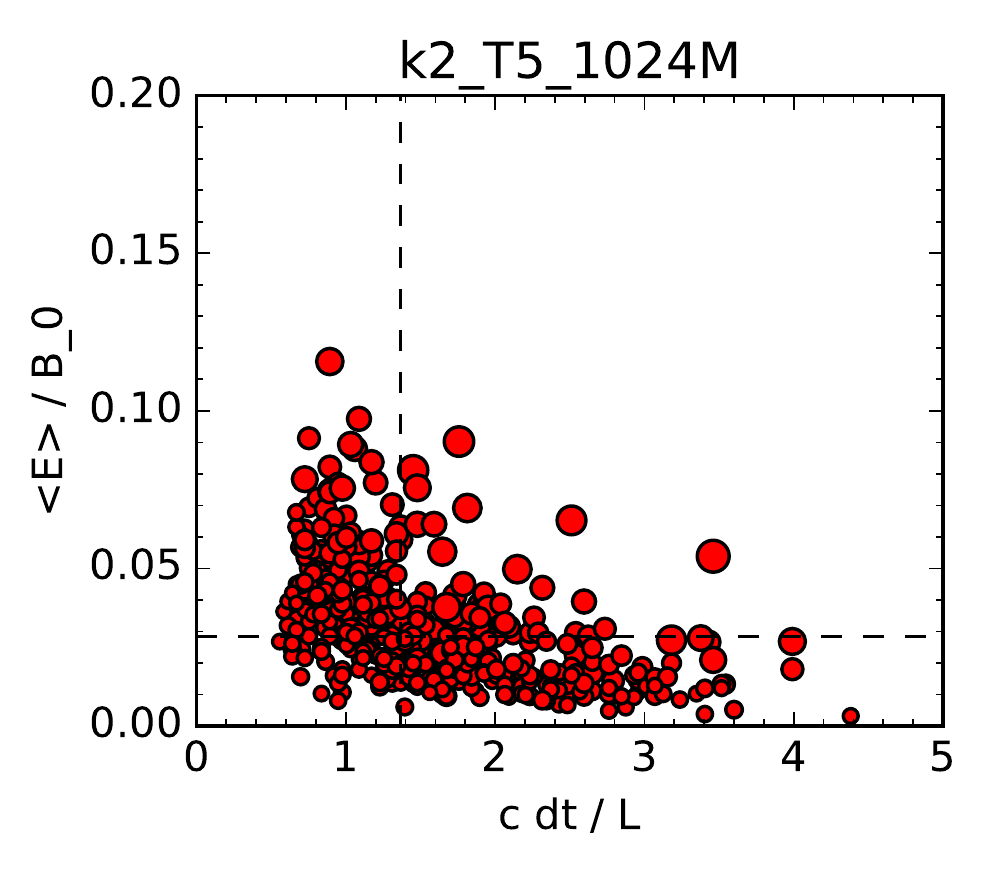}
\includegraphics[width=0.32\textwidth]{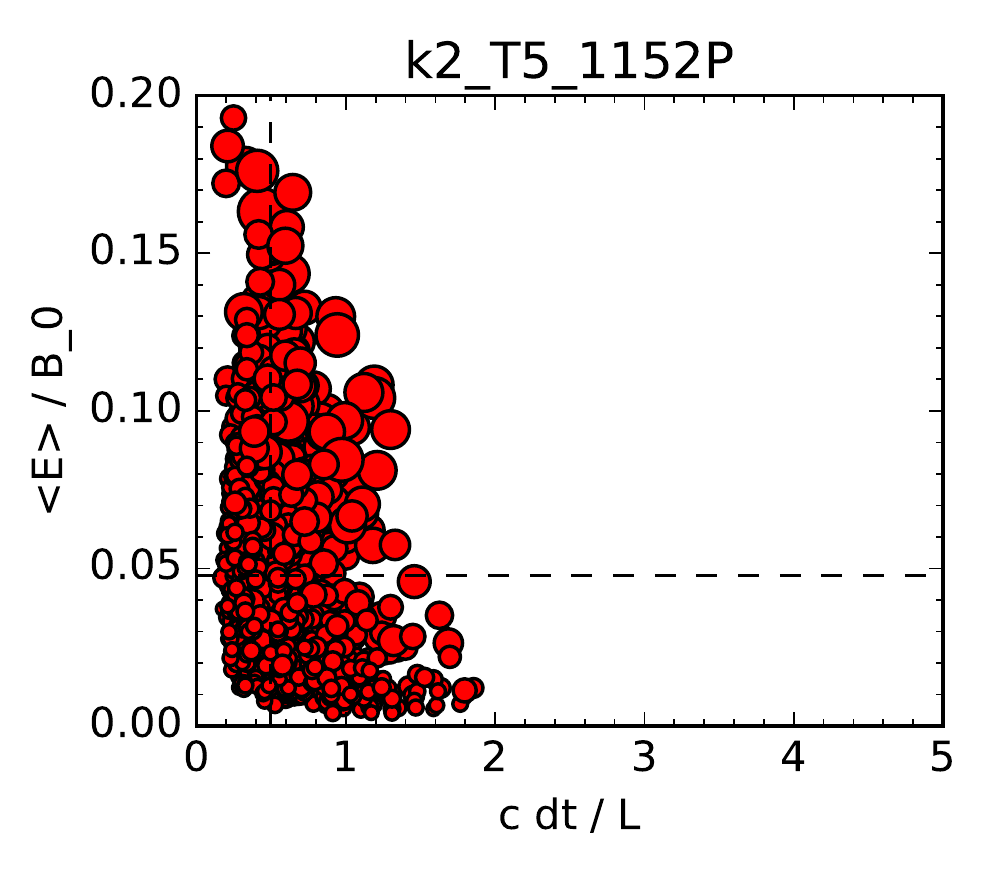}
\includegraphics[width=0.32\textwidth]{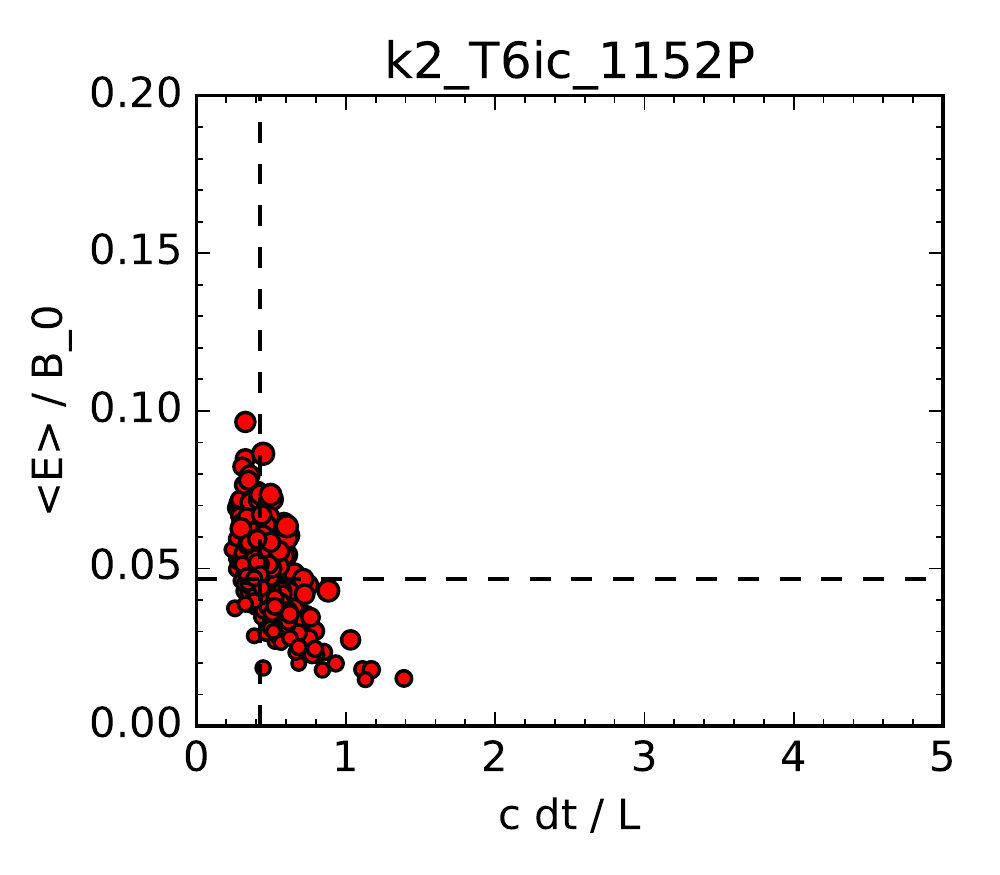}
\caption{Distribution of the time duration $c\Delta t/L$ vs. the effective acceleration electric field $\left<E_\parallel\right>_t/B_0 = \Delta\gamma(\rho_0/c\Delta t)$ for the main acceleration episodes of individually tracked particles compared for simulations k2\_T5\_1024M (left), k2\_T5\_1152P (middle), and k2\_T6ic\_1152P (right).}
\label{fig_orbits_stat2}
\end{figure*}

Figure \ref{fig_orbits_stat2} shows the distributions of time duration $\Delta t = t_2-t_1$ vs. effective acceleration electric field $\left<E_\parallel\right>_t / B_0$ for the MAE of all tracked energetic particles compared for 3 simulations. The maximum particle energy $\gamma_{\rm max}$ (not necessarily obtained at the end of the MAE) is indicated by the symbol size. For simulation k2\_T5\_1024M, initiated without an explicit magnetic perturbation, time durations of MAE are typically longer than $L/c$, with the median value of $\simeq 1.4(L/c)$. At the same time, the effective electric field is less than $0.05B_0$, with the median value of $\simeq 0.03B_0$. On the other hand, for simulation k2\_T5\_1152P the durations of MAE are typically shorter than $L/c$, with the median value of $\simeq 0.5(L/c)$, for a median electric field of $\simeq 0.048B_0$. Similar median values are found for simulation k2\_T6ic\_1152P, despite a much lower number of energetic electrons in the overall tracked sample.

\section{Discussion}
\label{sec_disc}

\subsection{Comparison with 2D results}

Two-dimensional PIC simulations of ABC fields \cite{Nal16,Yua16} used the configuration given by Eq. (\ref{eq_abc3d}) without the two terms dependent on $z$ (and using coordinates $x',y'$ rotated by $45^\circ$). In that case, magnetic dissipation was dominated by symmetric collapsing X-points (magnetic nulls) accompanied by head-on collision of magnetic domains with the same sign of the out-of-plane $B_z$ component. That resulted in a guide-field reconnection with significant $\bm{E}\cdot\bm{B} \simeq E_zB_z$.

In three dimensions, we observe qualitative differences with respect to the 2D simulations. Magnetic dissipation occurs mainly along the magnetic minima. We find in our analysis of individual particle acceleration histories that $\bm{E}\cdot\bm{B}/B$ is not an effective measure of electric field accelerating these particles (as it appeared to be in 2D). We find that energetic particles can be accelerated at arbitrary magnetic inclination (even under strong synchrotron cooling), and that they are accelerated by electric fields that are roughly perpendicular to the local magnetic fields.
The effective acceleration electric fields $E_\parallel < 0.2B_0$ are comparable with the 2D study.
The time durations of the main acceleration periods are sensitive to the initial magnetic perturbation.
Without explicit perturbation (the case of simulation k2\_T5\_1024M), the acceleration times are found to be longer in 3D.
We should note, however, that the time sampling and smoothing of individual particle histories was performed slightly differently in the 2D study.

We do not confirm that the current layers forming in the linear stage of coalescence instability have a double perpendicular structure, with two thickness scales of perpendicular profile of current density $\bm{j}$ derived from a shorter scale of the particle density $n$ and a longer scale of the velocity field $\bm{v}$ (or $\bm{E}\cdot\bm{B}$), as suggested by the 2D simulations \citep{Nal16}.
However, we do find that structures in the velocity fields of individual particle species (electrons/positrons) are significantly broader than the density structures.
It is possible that a double structure of current layers could be revealed at higher numerical resolutions.

Basic parameters reported in Table \ref{tab_param} are comparable with those obtained in the 2D study of \cite{Nal16}. The exponential growth rates of coalescence instability scale with the mean hot magnetization: for low magnetizations $\sigma_{\rm hot} \simeq 0.8$, we obtain $\tau_{\rm E} \simeq 0.32$ in 3D vs. $\tau_{\rm E} \simeq 0.35$ in 2D; for high magnetizations $\sigma_{\rm hot} \simeq 6-7$, we obtain $\tau_{\rm E} \simeq 0.16$ in 3D vs. $\tau_{\rm E} \simeq 0.18$ in 2D. Non-thermal particle acceleration can be compared between the slow-cooling 3D simulation k2\_T5\_1152P with $\sigma_{\rm hot} \simeq 7$, $p \simeq 3$, $\gamma_{\rm max} \sim 200\Theta$ and $f_{\rm e} \simeq 0.3$; and the 2D simulation s55L1600 with $\sigma_{\rm hot} \simeq 5.5$, $p \simeq 2.5$, $\gamma_{\rm max} \sim 800\Theta$ and $f_{\rm e} \simeq 0.6$. Non-thermal acceleration at comparable magnetization levels appears to be more efficient in 2D.

\subsection{Magnetic relaxation efficiency}

We have seen in Section \ref{sec_res_totene} that in our simulations magnetic energy is dissipated with efficiencies ranging from $(25-32)\%$ for our $k_{\rm ini}=2$ simulations to $70\%$ for simulation k4\_T5\_1152P (see Table \ref{tab_param}).
We have also identified in Section \ref{sec_res_voldist} the dominant Fourier modes of the $B_x$ distribution in the final states of three simulations: a mixed mode $(1,1,0)$ for the $k_{\rm ini}=2$ simulations, and the $k=1$ ``ABC'' modes $(0,1,0)$ and $(0,0,1)$ for the $k_{\rm ini}=4$ simulation (see Figure \ref{fig_psd_modes}).
We can now use the Taylor relaxation theorem \citep{Tay74} which states that in the presence of magnetic reconnection (local departures from ideal MHD), high-order magnetic configurations should relax by inverse cascade to a Taylor state, satisfying the conservation of total magnetic helicity $\left<H\right>$.
Conservation of total magnetic helicity has been demonstrated in Section \ref{sec_res_totene}, although it is relatively poor ($3-10\%$) for simulations k4\_T5\_1152P and k2\_T5\_1024M (see Figure \ref{fig_tot_ene}).
Let us now assume that the Taylor state is $(1,1,0)$ for $k_{\rm ini} = 2$ and $(0,1,0)+(0,0,1)$ for $k_{\rm ini} = 4$, so we can estimate the theoretical magnetic dissipation efficiencies.
For the $k=1$ ``ABC'' mode $(0,1,0)+(0,0,1)$, this implies that $\left<B^2\right>_{\rm fin} = \left<B^2\right>_{\rm ini}/k_{\rm ini}$, and hence that $f_{\rm B} = 1-(1/k_{\rm ini})$. For $k_{\rm ini} = 4$, we would expect $f_{\rm B} = 0.75$, and hence our simulation k4\_T5\_1152P is relaxed in $\simeq 93\%$.

Further investigation of the mixed mode reveals that it is not isotropic, it can be exemplified by $B_x = -B_y = -B_1\cos(\alpha_1x+\alpha_1y)$ and $B_z = -B_3\sin(\alpha_1x+\alpha_1y)$ with an amplitude ratio of $b = B_3/B_1 \simeq 1.5$.
We then find that $A_{x(y)} = (B_1/B_3)B_{x(y)}/\alpha_1$ and $A_z = (B_3/2B_1)B_z/\alpha_1$, that $\left<H\right> = B_1^3/(\alpha_1B_3) + B_3^3/(4\alpha_1B_1)$. Applying the conservation of magnetic helicity $\left<H\right> = \left<B_{\rm ini}^2\right>/(k_{\rm ini}\alpha_1)$, we find that $\left<B_{\rm fin}^2\right> = B_1^2 + B_3^2/2 = 2b(2+b^2)/(4+b^4)\left<B_{\rm ini}^2\right>/k_{\rm ini}$. For $k_{\rm ini} = 2$ and $b \simeq 1.5$, we find $f_{\rm B} \simeq 30\%$. Compared with this value, our simulations k2\_T5\_1152P, k2\_T6\_1152P and k2\_T6ic\_1152P would be relaxed in $\simeq 85\%$. If, however, our $k=2$ simulations should eventually relax to the $k=1$ ``ABC'' state (as indicated for simulation k2\_T5\_1024M), corresponding to $f_{\rm B} = 0.5$, they have achieved only $(50-64)\%$ of theoretical magnetic relaxation.

\subsection{Magnetic energy cascade}

We note that \cite{ZraEas16} performed MHD and force-free simulations of 2D and 3D high-order ($k_{\rm ini} \gg 1$) ABC fields and found the magnetic energy spectrum of freely decaying turbulence to be consistent with the Kolmogorov scaling $\mathcal{E}_{B,k} \propto k^{-5/3}$, as expected for perpendicular modes of MHD turbulence \citep{Zhd18}. On the other hand, performing PIC simulations for $k_{\rm ini} \le 4$, we found magnetic energy cascades to be significantly steeper, roughly consistent with $\mathcal{E}_{B,k} \propto k^{-2.5}$. We suggest that the main reason for this is that our simulations did not develope a volume-filling forward magnetic cascade, as the short modes can be associated mainly with the localized kinetically thin current layers, and that the decay of magnetic energy in the inertial subrange requires significantly more time.

\subsection{Collapsing magnetic minima}

We have described in Section \ref{sec_res_maps} (see Figures \ref{fig_tot_maps} and \ref{fig_prof1d}, as well as Supplementary Movies \ref{movie_volume}--\ref{movie_part_ang}) a qualitatively novel mode of localized magnetic dissipation, which is referred to as the collapse of local magnetic minimum. This involves formation of a pair of current layers connected by common magnetic flux, and their dynamical interaction allowed by shifting magnetic domains. We suggest that this represents a generic (not specific to the highly symmetric ABC fields) mechanism of spontaneous magnetic dissipation within a complex network of unbound (not attached to a hard surface, as in the case of the solar corona) magnetic minima. We propose that such a mechanism could be applied to the specific problem of gamma-ray flares from the Crab Nebula \citep{Zra17,Lyu18} and for the rapid gamma-ray flares of blazars \citep{Aha07,Nal12,Ack16}. This scenario is somewhat reminiscent of the interaction of two current layers within the equatorial striped wind proposed by \cite{Bat13}, however in that case the interaction was enabled by emergence of large-scale tearing modes.

\section{Conclusions}
\label{sec_conc}

We presented the results of the first 3D PIC simulations of ABC fields for $k_{\rm ini} = 2,4$, and we characterize in detail the growth and saturation of ideal coalescence instability, formation of kinetically-thin current layers, localized magnetic dissipation and associated non-thermal particle acceleration, decaying turbulent cascade of magnetic energy, and the dominant magnetic modes of the final state.
We describe a novel scenario of localized magnetic dissipation involving a collapse of magnetic minima and dynamical merger of a pair of current layers, which can result in production of rapid flares of high-energy radiation, e.g., the gamma-ray flares of blazars and of the Crab Nebula.
Magnetic relaxation to the Taylor state (ABC field for $k=1$) is demonstrated for the case of $k_{\rm ini} = 4$.
Accelerated particles form power-law components of index as hard as $-2.3$ for effective hot magnetization of $\left<\sigma_{\rm hot}\right> \sim 6$.
These particles are accelerated by electric fields that are largely perpendicular to the local magnetic fields.

\section*{Supplementary Movies}
\begin{figure*}
\includegraphics[width=0.245\textwidth]
{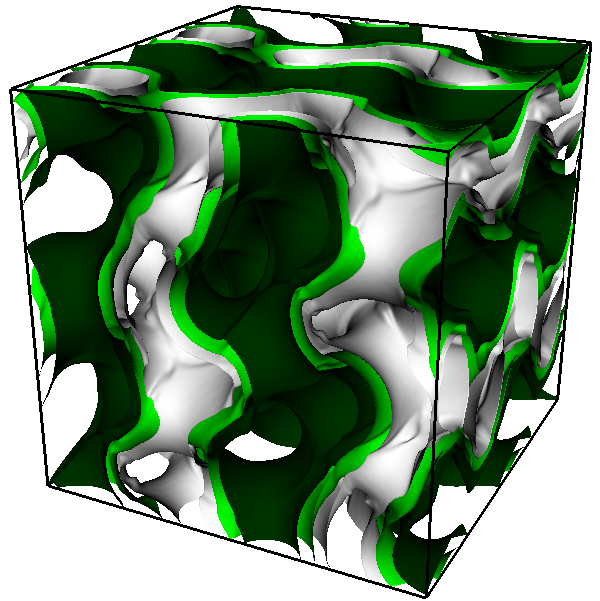}
\includegraphics[width=0.245\textwidth]
{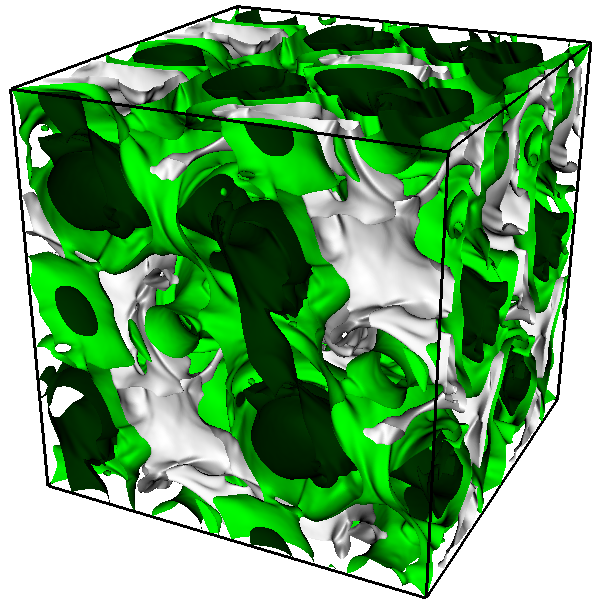}
\includegraphics[width=0.245\textwidth]
{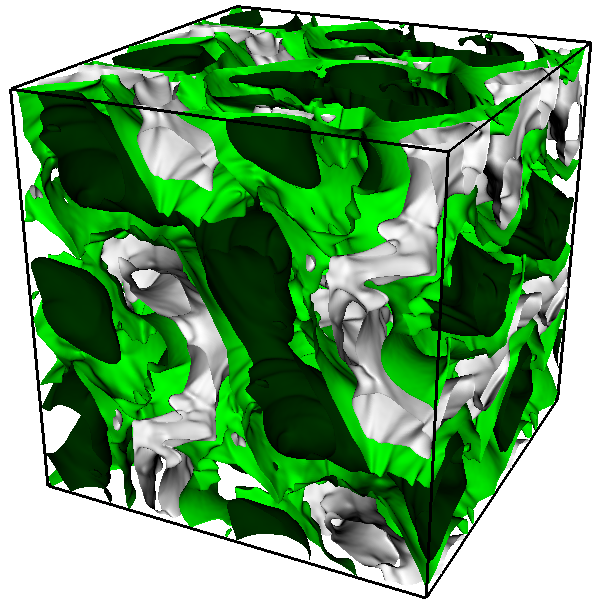}
\includegraphics[width=0.245\textwidth]
{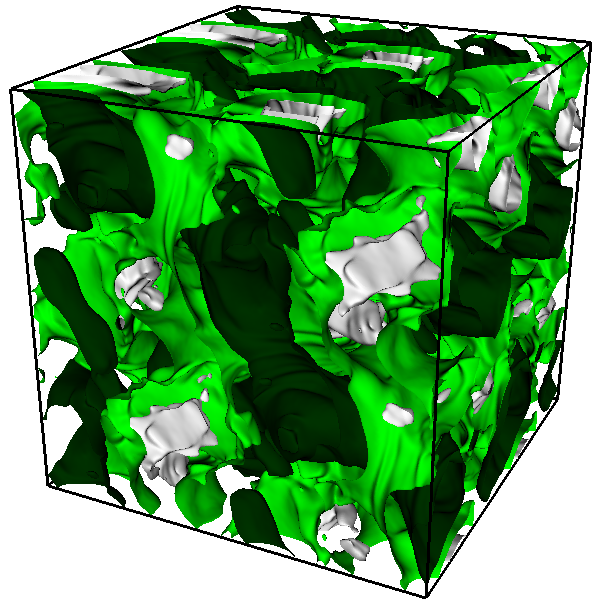}
\\
\includegraphics[width=0.245\textwidth]
{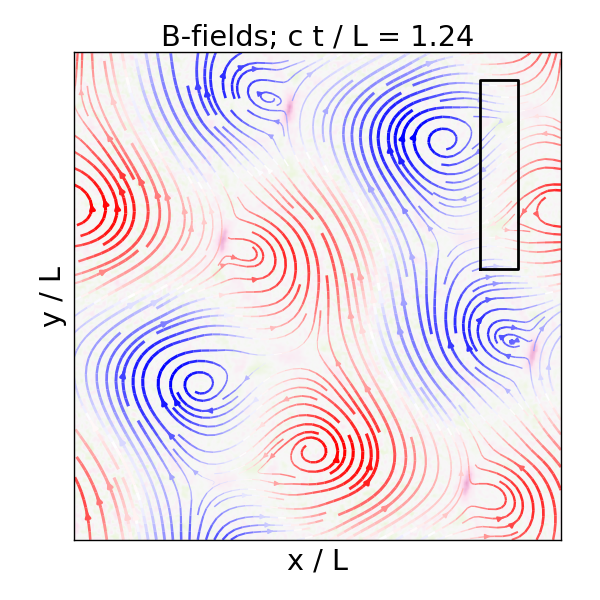}
\includegraphics[width=0.245\textwidth]
{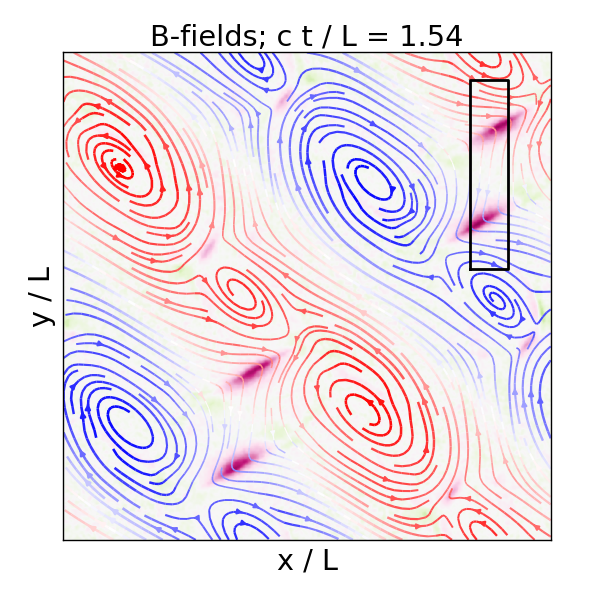}
\includegraphics[width=0.245\textwidth]
{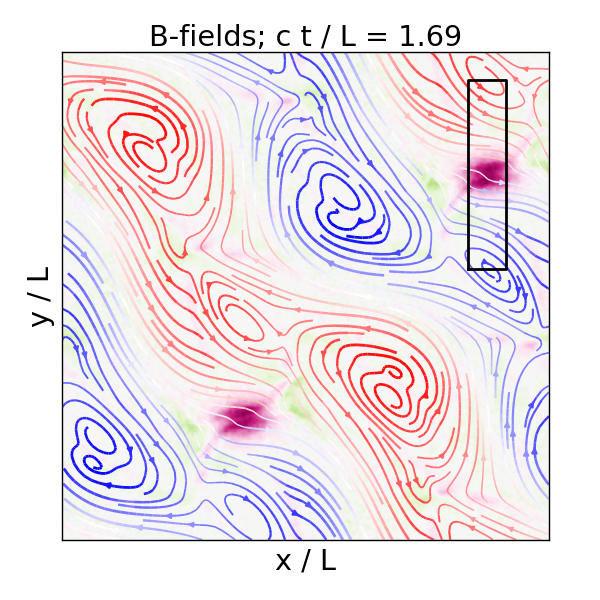}
\includegraphics[width=0.245\textwidth]
{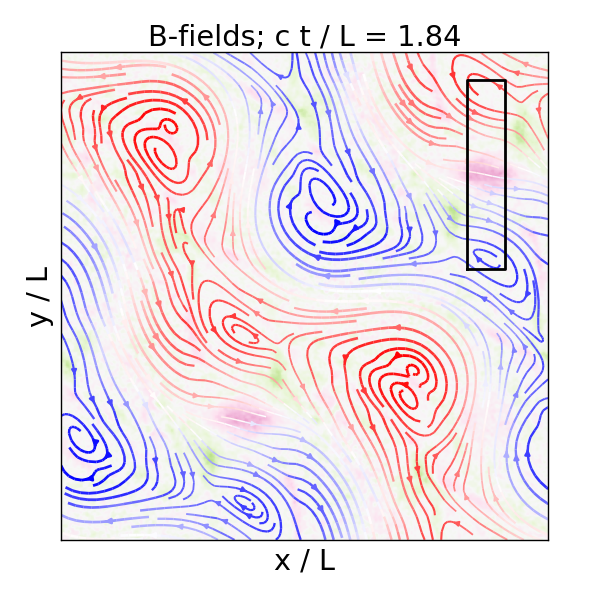}
\\
\includegraphics[width=0.245\textwidth]
{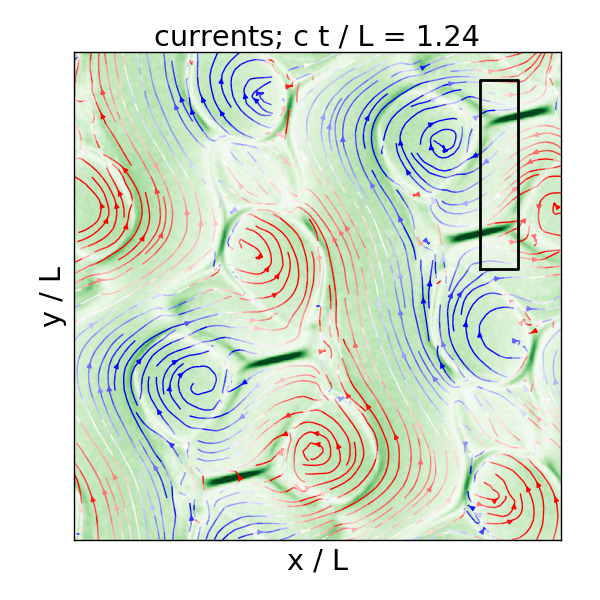}
\includegraphics[width=0.245\textwidth]
{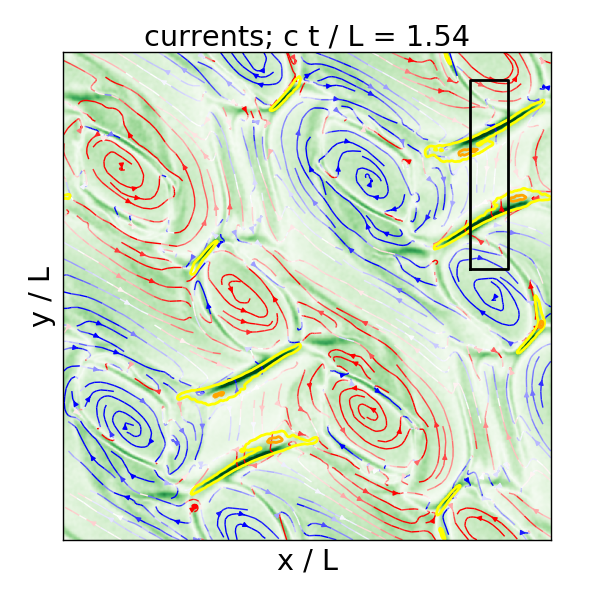}
\includegraphics[width=0.245\textwidth]
{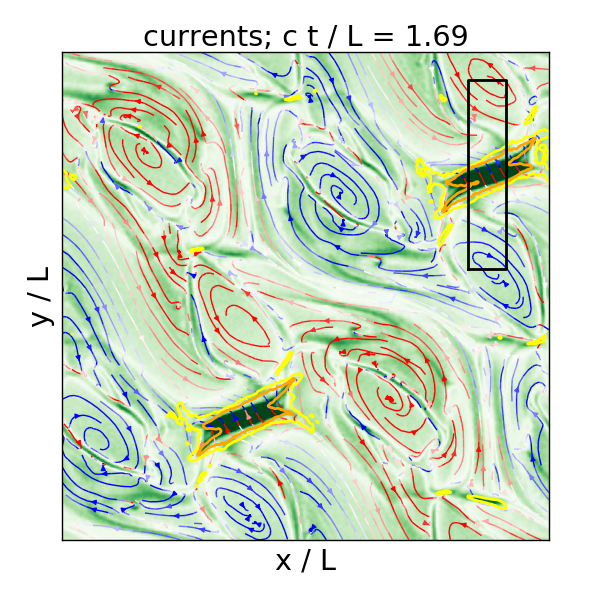}
\includegraphics[width=0.245\textwidth]
{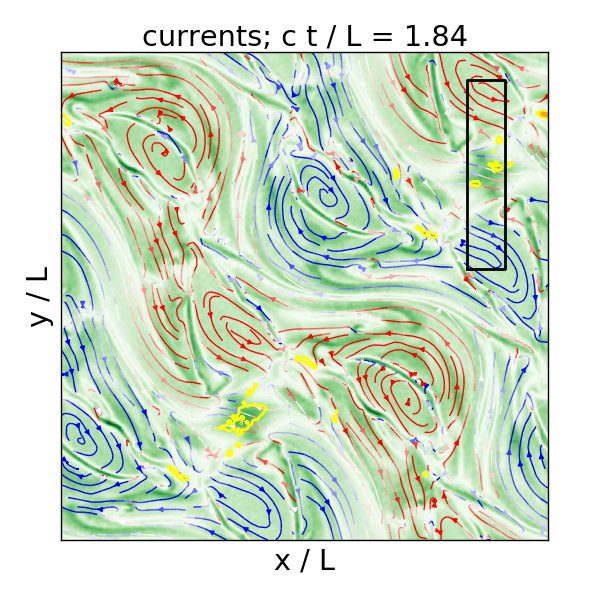}
\\
\includegraphics[width=0.245\textwidth]
{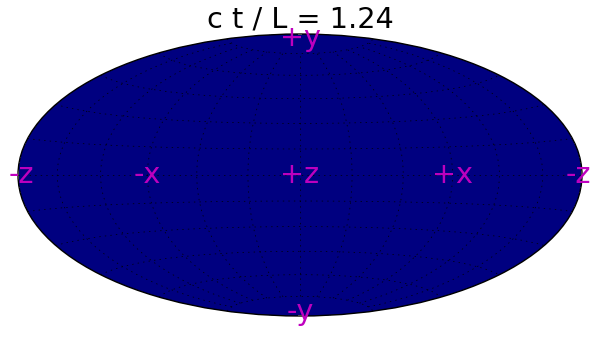}
\includegraphics[width=0.245\textwidth]
{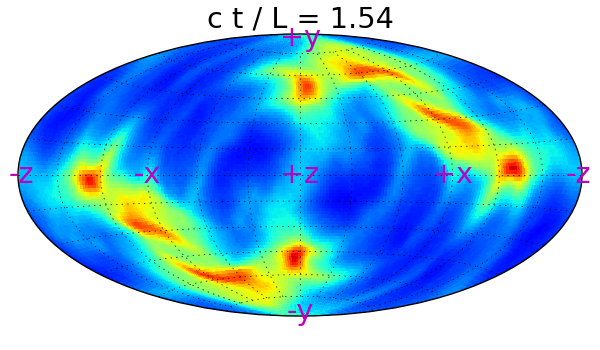}
\includegraphics[width=0.245\textwidth]
{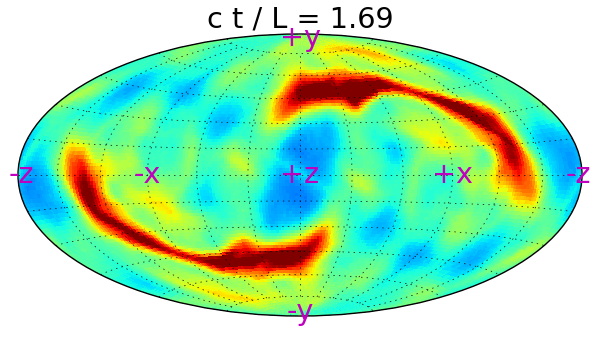}
\includegraphics[width=0.245\textwidth]
{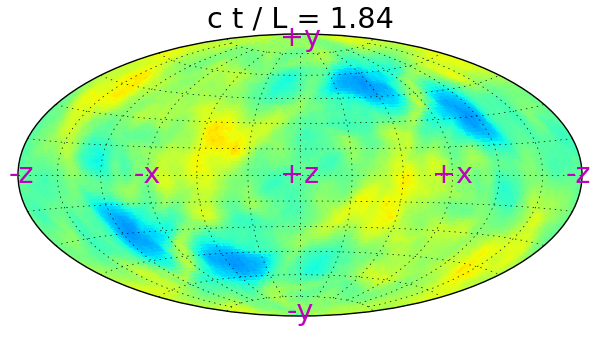}
\caption{Evolution of magnetic fields and currents during the main magnetic dissipation phase of simulation k2\_T5\_1152P (see Supplementary Movies \ref{movie_volume}--\ref{movie_part_ang}).
\emph{The top row of panels} presents volume rendering of the magnetic field strength (white - $B = 1.2B_0$, light green - $B = 1.5B_0$, dark green - $B = 1.8B_0$) for 4 simulation times $ct/L = 1.24,1.54,1.69,1.84$ (indicated also in Figure \ref{fig_tot_ene}). \emph{The second row of panels} presents the $(x,y)$ surface distribution of magnetic fields corresponding to the front face of the cube shown above ($z = 0$). The coordinate range is $x,y \in [0:L] = [0:900\rho_0]$. Here, the in-plane magnetic field orientation $(B_x,B_y)$ is indicated with the streamlines of arbitrary separation, the value of out-of-plane field $B_z$ is indicated with streamline color (red - positive, blue - negative), and the magnetic field strength $|\bm{B}|$ is indicated with streamline thickness. The purple/green color patches indicate the negative/positive values of $\bm{E}\cdot\bm{B}$. The black box indicates Patch A, from which we extract the $y$ profiles of plasma parameters shown in Figure \ref{fig_prof1d}. \emph{The third row of panels} presents the $(x,y)$ surface distribution of current densities on the $z = 0$ surface. Here, the in-plane current density $(j_x,j_y)$ is indicated with the streamlines, the value of out-of-plane current density $j_z$ is indicated with the streamline color (red - positive, blue - negative), and the current density magnitude $|\bm{j}|$ is indicated with the color shading (dark green patches indicate the most intense current density). The thick yellow/orange contours indicate the regions of hot plasma with mean particle energy $\left<\gamma\right>/\Theta = 10,15$, respectively. \emph{The bottom row of panels} presents the angular distribution of energetic electrons and positrons with momentum $u > 40\Theta$ for the same 4 simulation times.}
\label{fig_tot_maps}
\end{figure*}

\begin{figure*}
\includegraphics[width=\textwidth]{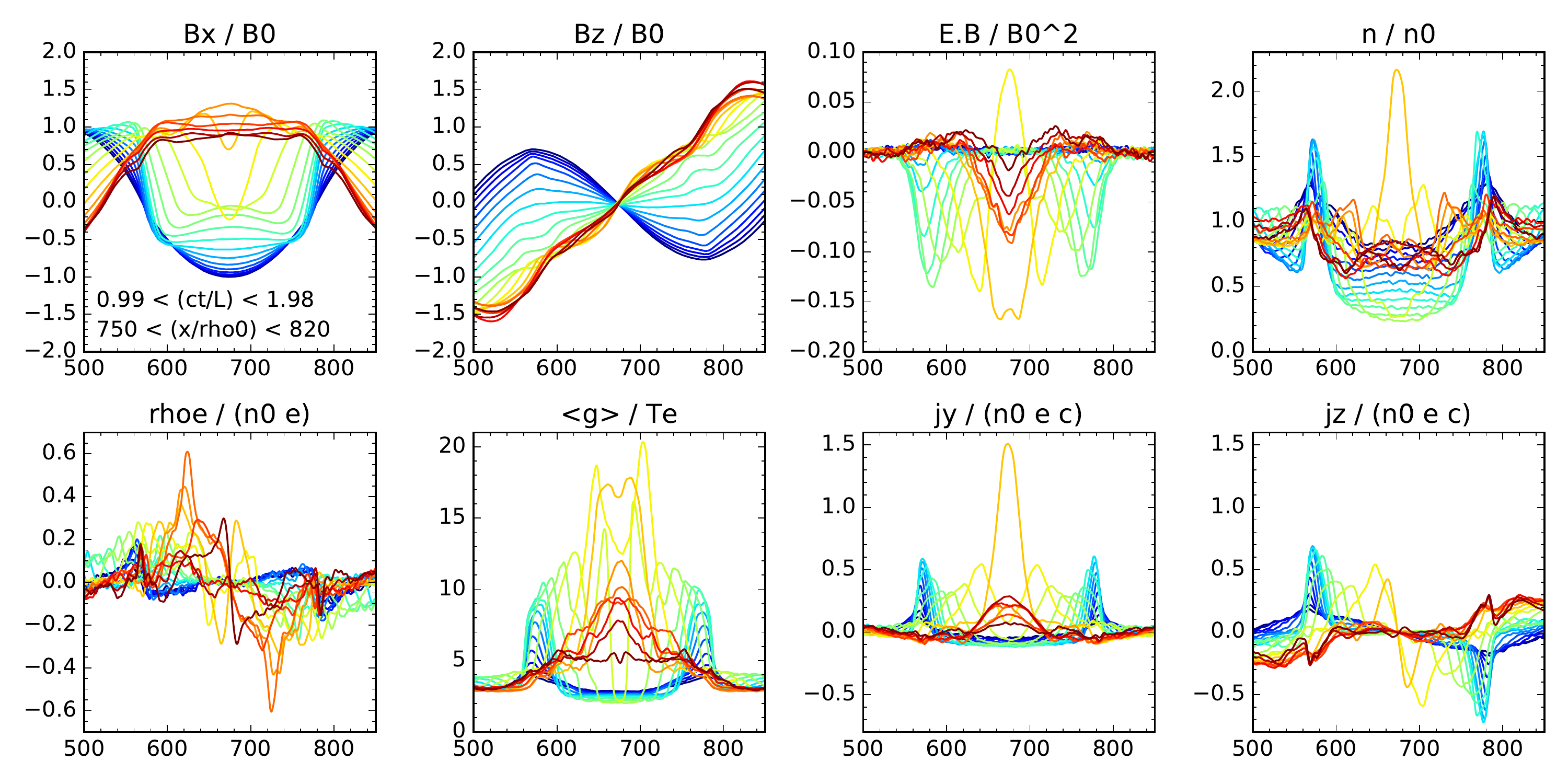}
\caption{Plasma parameters extracted from the Patch A of the $z = 0$ surface, defined by $750 < x/\rho_0 < 820$ and $500 < y/\rho_0 < 850$ (indicated with gray boxes in the middle row of panels in Figure \ref{fig_tot_maps}). The data are averaged over the $x$ coordinate and presented as functions of the $y$ coordinate. Line colors indicate the progression of simulation time from deep blue to brown over the range $0.99 < ct/L < 1.98$. From the top left, the panels show the following dimensionless parameters: magnetic field components $B_x/B_0$ and $B_z/B_0$; non-ideal field scalar $(\bm{E}\cdot\bm{B})/B_0^2$; particle number density $n/n_0$; charge density $\rho_{\rm e} / (en_0)$; mean particle energy $\left<\gamma\right>/\Theta$; and current density components $j_y/(ecn_0)$ and $j_z/(ecn_0)$.}
\label{fig_prof1d}
\end{figure*}

The following movies are available at \url{http://users.camk.edu.pl/knalew/abc3d/}:
\begin{enumerate}
\item
\label{movie_volume}
Volume rendering of $B$ and $\bm{E}\cdot\bm{B}$ (Figure \ref{fig_tot_maps});
\item
\label{movie_xymap}
Sections of magnetic fields and currents at the $z=0$ surface (Figure \ref{fig_tot_maps});
\item
\label{movie_part_ang}
Angular distribution of energetic electrons and positrons (Figure \ref{fig_tot_maps});
\item
\label{movie_orbits}
Acceleration histories of individual tracked particles (Figure \ref{fig_orbit_example}).
\end{enumerate}

\section*{Acknowledgments}
We thank the reviewer -- Maxim Lyutikov -- for helpful suggestions.
We acknowledge discussions with Jonathan Zrake, Yajie Yuan, Varadarajan Parthasarathy and Jos{\'e} Ortu{\~n}o Mac{\'i}as.
The original version of the {\tt Zeltron} code was created by Beno{\^i}t Cerutti and co-developed by Gregory Werner at the University of Colorado Boulder (\url{http://benoit.cerutti.free.fr/Zeltron/}).
These results are based on numerical simulations performed at supercomputers: \emph{Mira} at Argonne National Laboratory, USA (2016 INCITE allocation; PI: D. Uzdensky); and \emph{Prometheus} at Cyfronet AGH, Poland (PLGrid grant {\tt abcpic17}; PI: K. Nalewajko).
This work was supported by the Polish National Science Centre grant 2015/18/E/ST9/00580.



\begin{table*}
\begin{center}
\caption{Basic parameters of the 3D simulations described in this work: $N_c$ is the number of numerical grid cells; $L/\rho_0$ is the physical size of simulation domain in units of the nominal gyroradius $\rho_0 = \Theta m_{\rm e}c^2/(eB_0)$; $\Theta = k_{\rm B}T/(m_{\rm e}c^2)$ is the relativistic temperature of the initial particle energy distribution; $\tilde{a}_1$ is the normalization constant for the dipole moments of the local particle angular distributions; $\left<\sigma_{\rm hot}\right>_{\rm ini}$ is the mean hot magnetization value defined by Eq. (\ref{eq_sigma_hot}); $\tau_{\rm E}$ is the peak exponential growth rate of the total electric energy in units of $ct/L$; $\beta_{\rm rec} = [\left<E^2\right>_{\rm peak} / \left<B^2\right>_{\rm ini}]^{1/2}$ is the effective reconnection rate; $f_{\rm B} = 1-\left<B^2\right>_{\rm fin}/\left<B^2\right>_{\rm ini}$ is the efficiency of magnetic dissipation; $f_n$ and $f_e$ are the peak values of the number and energy fractions of the non-thermal high-momentum excess in the particle momentum distribution; $p$ is the power-law index of the particle momentum distribution $N(u) \propto u^{-p}$; and $\gamma_{\rm max}/\Theta$ is the peak value of the maximum particle energy evaluated at the $10^{-3}$ level of the normalized $u^2 N(u)$ distribution.}
\label{tab_param}
\begin{tabular}{lccccccccccccc}
\hline
name & $N_c$ & $L/\rho_0$ & $\Theta$ & $\tilde{a}_1$ & $\left<\sigma_{\rm hot}\right>_{\rm ini}$ & $\tau_{\rm E}$ & $\beta_{\rm rec}$ & $f_{\rm B}$ & $f_n$ & $f_e$ & $p$ & $\gamma_{\rm max}/\Theta$ & {\rm remarks} \\
\hline
k2\_T5\_1024M   & $1024^3$ & 400 & $10^5$ & 0.4 & 3.2 & 0.213 & 0.24 & 0.32 & 0.05 & 0.16 & 3.9 & 90 & {\rm no\,\,perturbation} \\
k2\_T5\_1152P   & $1152^3$ & 900 & $10^5$ & 0.4 & 7.2 & 0.161 & 0.31 & 0.25 & 0.09 & 0.31 & 3.0 & 200 \\
k2\_T6\_1152P   & $1152^3$ & 900 & $10^6$ & 0.2 & 3.6 & 0.195 & 0.26 & 0.26 & 0.11 & 0.31 & 2.4 & 60 \\
k2\_T6ic\_1152P & $1152^3$ & 900 & $10^6$ & 0.2 & 3.6 & 0.189 & 0.27 & 0.26 & 0.09 & 0.29 & 2.3 & 50 & {\rm SYN+IC\,\,cooling} \\ 
k4\_T5\_1152P   & $1152^3$ & 450 & $10^5$ & 0.2 & 0.9 & 0.315 & 0.15 & 0.70 & 0.03 & 0.09 & --  & 40 & {\rm no\,\,perturbation} \\
\hline
\end{tabular}
\end{center}
\end{table*}


\begin{figure*}
\includegraphics[width=\textwidth]{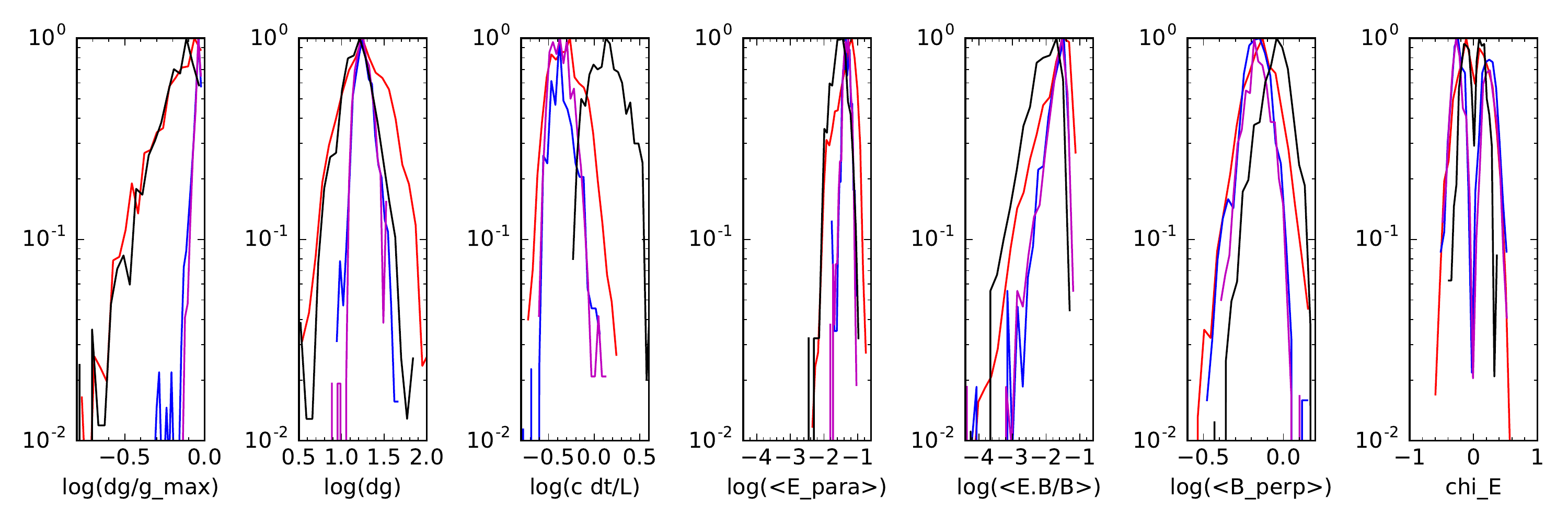}
\caption{Statistics of the main acceleration episodes (MAE) for all tracked energetic particles compared for all simulations with $k_{\rm ini} = 2$. From the left, the panels show the normalized log-histograms of: (1) the ratio of energy gain during the MAE $\Delta\gamma = \gamma(t_2)-\gamma(t_1)$ to the maximum particle energy $\gamma_{\rm max}$ (which may be obtained outside of MAE); (2) the value of $\Delta\gamma$ normalized to initial particle temperature $\Theta$; (3) the time duration of the MAE $\Delta t = t_2-t_1$, normalized to $L/c$; (4) the effective acceleration electric field calculated as $\left<E_\parallel\right>_t/B_0 = \Delta\gamma(\rho_0/c\Delta t)$; (5) the mean electric field component parallel to the magnetic field $\left<\bm{E}\cdot(\bm{B}/B)\right>_t/B_0$; (6) the mean value of perpendicular magnetic field $\left<|\bm{B}\times(\bm{u}/u)|\right>_t/B_0$; (7) the mean angle between particle momentum and electric field $\chi_{u,E} = (\bm{u}/u)\cdot(\bm{E}/E)$. The line styles are the same as in Figure \ref{fig_tot_ene}.}
\label{fig_orbits_stat1}
\end{figure*}

\appendix
\section{Linear analysis of the dominant coalescence mode}
\label{app1}

\emph{(This appendix is modified with respect to the accepted version of the manuscript.)}

Here we present an analytical estimate of the growth rate of the dominant coalescence mode in the framework of ideal MHD. We adopt the following notation: background magnetic field $\bm{B}_0$ as given by Eq. (\ref{eq_abc3d}), background current density $\bm{j}_0 = (\alpha_kc/4\pi)\bm{B}_0$, perturbation magnetic field $\bm{B}_1 \propto \exp(\omega t)$ as given by Eq. (\ref{eq_Bpert}), and perturbation electric field $\bm{E}_1$.
This particular perturbation satisfies the Beltrami conditions $\bm\nabla\times\bm{B}_1 = (\alpha_k/\sqrt{2})\bm{B}_1$ and $\bm\nabla\times\bm{E}_1 = (\alpha_k/\sqrt{2})\bm{E}_1$.
Using the linearized Maxwell's equations:
\begin{eqnarray}
\partial_t\bm{B}_1 &=& -c\bm\nabla\times\bm{E}_1\,,
\\
\partial_t\bm{E}_1 &=& c\bm\nabla\times\bm{B}_1 - 4\pi\bm{j}_1\,,
\end{eqnarray}
we find direct expressions for $\bm{E}_1$ and $\bm{j}_1$:
\begin{eqnarray}
\bm{E}_1 &=& -\left(\frac{\sqrt{2}\omega}{\alpha_kc}\right)\bm{B}_1\,,
\\
\bm{j}_1 &=& \left(\frac{2\omega^2+\alpha_k^2c^2}{4\pi\sqrt{2}\alpha_kc}\right)\bm{B}_1\,.
\end{eqnarray}
Now we demand that the linearized $\bm{j}\times\bm{B}$ force vanishes, otherwise a residual force would drive a velocity field that in turn would generate a divergent series of additional electromagnetic modes:
\begin{eqnarray}
\bm{j}_0\times\bm{B}_1 + \bm{j}_1\times\bm{B}_0 &=& 0\,,
\\
\frac{(\sqrt{2}-1)\alpha_k^2c^2-2\omega^2}{4\pi\sqrt{2}\alpha_kc}(\bm{B}_0\times\bm{B}_1) &=& 0\,.
\end{eqnarray}
This yields a unique value of $\omega$:
\begin{eqnarray}
\frac{\omega}{\alpha_kc} &=& \left(\frac{\sqrt{2}-1}{2}\right)^{1/2} \simeq 0.455\,,
\end{eqnarray}
which corresponds to the growth rate discussed in Section \ref{sec_res_totene}: $\tau_E = c/(2\omega L) = (4\pi k_{\rm ini}\omega/\alpha_kc)^{-1} \simeq 0.09$ for $k_{\rm ini} = 2$, that is somewhat shorter than measured in the simulations.


\begin{thebibliography}{}

\bibitem[{Abdo} {et~al.}(2011)]{Abd11}
{Abdo}, A.~A., {Ackermann}, M., {Ajello}, M., {et~al.}, 2011, Science, 331, 739

\bibitem[Ackermann et~al.(2016)]{Ack16}
Ackermann, M., Anantua, R., Asano, K., et~al.\ 2016, ApJ, 824, L20

\bibitem[{Aharonian} {et~al.}(2007)]{Aha07}
{Aharonian}, {F.}, {et~al.}, 2007, ApJ, 664, L71

\bibitem[Baty et~al.(2013)]{Bat13}
Baty, H., Petri, J., \& Zenitani, S.\ 2013, MNRAS, 436, L20

\bibitem[{Begelman} {et~al.}(1984)]{Beg84}
{Begelman}, {M.~C.}, {Blandford}, {R.~D.}, {Rees}, M.~J., 1984, RvMP, 56, 255

\bibitem[{Blandford} \& {Znajek}(1977)]{Bla77}
{Blandford}, R.~D., \& {Znajek}, R.~L., 1977, MNRAS, 179, 433

\bibitem[Blandford et~al.(2015)]{Bla15}
Blandford, R., East, W., Nalewajko, K., Yuan, Y., \& Zrake, J.\ 2015, arXiv:1511.07515

\bibitem[Cerutti et~al.(2012)]{Cer12}
Cerutti, B., Werner, G.~R., Uzdensky, D.~A., \& Begelman, M.~C.\ 2012, ApJ, 754, L33

\bibitem[Cerutti et~al.(2013)]{Cer13}
Cerutti, B., Werner, G.~R., Uzdensky, D.~A., \& Begelman, M.~C.\ 2013, ApJ, 770, 147

\bibitem[Cerutti et~al.(2014)]{Cer14}
Cerutti, B., Werner, G.~R., Uzdensky, D.~A., \& Begelman, M.~C.\ 2014, ApJ, 782, 104

\bibitem[Coroniti(1990)]{Cor90}
Coroniti, F.~V.\ 1990, ApJ, 349, 538

\bibitem[Dombre et al.(1986)]{Dom86}
Dombre, T., Frisch, U., Henon, M., Greene, J.~M., \& Soward, A.~M.\ 1986, JFM, 167, 353

\bibitem[East et~al.(2015)]{Eas15}
East, W.~E., Zrake, J., Yuan, Y., \& Blandford, R.~D.\ 2015, PhRvL, 115, 095002

\bibitem[Esirkepov(2001)]{Esi01}
Esirkepov, T.~Z.\ 2001, CoPhC, 135, 144

\bibitem[Guo et~al.(2014)]{Guo14}
Guo, F., Li, H., Daughton, W., \& Liu, Y.-H.\ 2014, PhRvL, 113, 155005

\bibitem[Hoshino(2012)]{Hos12}
Hoshino, M.\ 2012, PhRvL, 108, 135003

\bibitem[{Komissarov} {et~al.}(2007)]{Kom07}
{Komissarov}, S.~S., {Barkov}, M.~V., {Vlahakis}, N., {K{\"o}nigl}, A., 2007, MNRAS, 380, 51

\bibitem[Li et~al.(1992)]{Li92}
Li, Z.-Y., Chiueh, T., \& Begelman, M.~C.\ 1992, ApJ, 394, 459

\bibitem[Lyubarsky \& Kirk(2001)]{Lyub01}
Lyubarsky, Y., \& Kirk, J.~G.\ 2001, ApJ, 547, 437

\bibitem[Lyutikov et~al.(2017)]{Lyu17}
Lyutikov, M., Sironi, L., Komissarov, S.~S., \& Porth, O.\ 2017, JPlPh, 83, 635830602

\bibitem[Lyutikov et~al.(2018)]{Lyu18}
Lyutikov, M., Komissarov, S., \& Sironi, L.\ 2018, JPlPh, 84, 635840201

\bibitem[Nalewajko et~al.(2012)]{Nal12}
Nalewajko, K., Begelman, M.~C., Cerutti, B., Uzdensky, D.~A., \& Sikora, M. 2012, MNRAS, 425, 2519

\bibitem[Nalewajko et al.(2016)]{Nal16}
Nalewajko, K., Zrake, J., Yuan, Y., East, W.~E., \& Blandford, R.~D.\ 2016, \apj, 826, 115

\bibitem[Nalewajko et~al.(2018)]{Nal18}
Nalewajko, K., Yuan, Y., \& Chru{\'s}li{\'n}ska, M.\ 2018, JPlPh, 84, 755840301

\bibitem[Sironi \& Spitkovsky(2014)]{Sir14}
Sironi, L., \& Spitkovsky, A.\ 2014, ApJ, 783, L21

\bibitem[Tavani et~al.(2011)]{Tav11}
Tavani, M., Bulgarelli, A., Vittorini, V., et~al.\ 2011, Science, 331, 736

\bibitem[{Taylor(1974)}]{Tay74}
Taylor, J.~B. 1974, PhRvL, 33, 1139

\bibitem[Vay(2008)]{Vay08}
Vay, J.-L.\ 2008, PhPl, 15, 056701

\bibitem[Werner et~al.(2016)]{Wer16}
Werner, G.~R., Uzdensky, D.~A., Cerutti, B., Nalewajko, K., \& Begelman, M.~C.\ 2016, ApJ, 816, L8

\bibitem[Yuan et~al.(2016)]{Yua16}
Yuan, Y., Nalewajko, K., Zrake, J., East, W.~E., \& Blandford, R.~D.\ 2016, ApJ, 828, 92

\bibitem[{Zenitani} \& {Hoshino}(2001)]{Zen01}
{Zenitani}, S., \& {Hoshino}, M., 2001, ApJ, 562, L63

\bibitem[Zhdankin et~al.(2018)]{Zhd18}
Zhdankin, V., Uzdensky, D.~A., Werner, G.~R., \& Begelman, M.~C.\ 2018, MNRAS, 474, 2514

\bibitem[Zrake \& East(2016)]{ZraEas16}
Zrake, J., \& East, W.~E.\ 2016, ApJ, 817, 89

\bibitem[Zrake \& Arons(2017)]{Zra17}
Zrake, J., \& Arons, J.\ 2017, ApJ, 847, 57

\end{thebibliography}
\end{document}